\documentclass[preprint,preprintnumbers,amsmath,amssymb]{revtex4-1}
\usepackage{amsfonts}    
\usepackage{amssymb}
\usepackage{amsmath}
\usepackage{latexsym}
\usepackage{eepic}
\usepackage{graphicx}
\usepackage{rotating}
\usepackage[usenames,dvipsnames]{color}
\usepackage{color}

\usepackage[active]{srcltx}

\newlength\tindent
\def\be {\begin {equation}}
\def\ee {\end {equation}}

\newcommand{\pa}{\partial}

\newcommand{\Om}{\Omega}
\newcommand{\om}{\omega}

\newcommand{\De}{\Delta}

\newcommand{\rar}{\rightarrow}

\newcommand{\non}{\nonumber}

\begin{document}

\title{Four-body (an)harmonic oscillator in $d$-dimensional space: $S$-states, (quasi)-exact-solvability,
hidden algebra $sl(7)$ }

\author{
           M.A.~Escobar-Ruiz,\\[8pt]
   Departamento de F\'isica, UAM-I, M\'exico DF 09340, M\'exico\\[6pt]
   admau@xanum.uam.mx\\[8pt]
Alexander V. Turbiner\\[8pt]
Instituto de Ciencias Nucleares, UNAM, M\'exico DF 04510, M\'exico \\[8pt]
turbiner@nucleares.unam.mx
 \\[10pt]
and \\[10pt]
           Willard Miller, Jr.\\[8pt]
School of Mathematics, University of Minnesota, \\
Minneapolis, Minnesota, U.S.A.\\[8pt]
miller@ima.umn.edu
}

\begin{abstract}
As a generalization and extension of our previous paper {\it J. Phys. A: Math. Theor. 53 055302} \cite{AME2020},
in this work we study a quantum 4-body system in $\mathbb{R}^d$ ($d\geq 3$) with quadratic and sextic pairwise potentials in the {\it relative distances}, $r_{ij} \equiv {|{\bf r}_i - {\bf r}_j |}$, between particles. Our study is restricted to solutions in the space of relative motion with zero total angular momentum ($S$-states). In variables $\rho_{ij} \equiv r_{ij}^2$, the corresponding reduced Hamiltonian of the system possesses a hidden $sl(7;{\bf R})$ Lie algebra structure. In the $\rho$-representation it is shown that the 4-body harmonic oscillator with arbitrary masses and unequal spring constants is exactly-solvable (ES). We pay special attention to the case of four equal masses and to atomic-like (where one mass is infinite, three others are equal), molecular two-center (two masses are infinite, two others are equal) and molecular three-center (three infinite masses) cases. In particular, exact results in the molecular case are compared with those obtained within the Born-Oppenheimer approximation. The first and second order symmetries of non-interacting system are searched. Also, the reduction to the lower dimensional cases $d=1,2$ is discussed. It is shown that for four body harmonic oscillator case there exists an infinite family of eigenfunctions which depend on the single variable which is the moment-of-inertia of the system.
\end{abstract}
\maketitle

\section{Introduction}

The problem of two point masses connected by a spring that obeys Hooke's
law is one of the basic problems in classical mechanics: it occurs in practically all textbooks. For any dimension $d\geq1$ this system is reduced to a one-dimensional simple oscillator which vibrates around its center-of-mass and for $d>1$ can rotate simultaneously. It can be easily quantized, it appears in the literature under the name of the harmonic oscillator and it is more than well-known. The situation changes dramatically when three or more arbitrary particles are connected through springs. In this case a complete separation of variables in the Hamiltonian is not possible and normal coordinates do not exist. In general, the collective motion can not be expressed as a linear superposition of decoupled simple oscillators. As the result a multidimensional oscillator motion is represented by different pendulums (e.g. Foucault or oscilloscope).

In quantum mechanics, the hyperspherical-harmonic expansion method introduced in 1935 by Zernike and Brinkman\cite{Zernike} and reintroduced 25 years later in a different form by Delves\cite{Delves:1958} and Smith\cite{Smith}, provides a very general approach to the $N$-body problem. In the case of four particles connected via harmonic-oscillator two-body forces such a method is useful only when the particles are identical and at the same time the spring constants are equal. In this case it provides the exact solution of the problem. However, it is very important to find exact solutions when masses are arbitrary and the spring constants are different as well.

In our previous paper \cite{AME2020} the quantum three-body oscillator in $d-$dimensions was fully analyzed in the case of arbitrary masses, unequal spring constants and zero total angular momentum ($S-$states). A reduced Hamiltonian of the system was constructed explicitly and it was shown that the problem is exactly-solvable. In the present consideration we move to the quantum four-body oscillator system, namely four arbitrary masses moving in $\mathbb{R}^d$ ($d>2$) with a quadratic pairwise harmonic potential in mutual (relative) distances variables $r_{ij}$. We will restrict ourselves to the case of zero total angular momentum ($S-$states).

The main goal of the present study is to show that this 4-body quantum oscillator can be treated and solved in a simpler way by using variables employed by the authors in an earlier series of papers \cite{Willard:2018}-\cite{MTE:2018}. The motivation of the present paper is three-fold. Our first aim is to reveal and describe in detail the underlying hidden $sl(7;R)$ Lie algebraic structure of the reduced Hamiltonian which explains the existence of exact analytical solutions.

Secondly, we will determine how this algebraic structure degenerates in the physically important atomic-like(where one mass is infinite), molecular-like two-center (two masses are infinite) and molecular-like three-center (three infinite masses) cases. In particular, for each of these systems it will be shown that the reduced problem is gauge-equivalent to a Schr\"{o}dinger equation in a non-conformally flat space. The explicit form of the corresponding Schr\"{o}dinger-operator is presented.

Thirdly, for any dimension $d\geq 1$ and arbitrary masses we will construct an exactly-solvable model that effectively depends on a single variable which is proportional to the moment-of-inertia of the system. It corresponds to a certain 4-body oscillator system, and it admits a nontrivial quasi-exactly-solvable extension.

The structure of the article is organized very simply. In Section \ref{generalities}, we first review the 4-body \emph{Jacobi harmonic oscillator} which is an exactly and quasi-exactly-solvable system. Next, in Section \ref{4bodyS} we describe for an arbitrary potential $V=V(\rho_{ij})$, $\rho_{ij}=r_{ij}^2$, the reduction of the quantum four-body Hamiltonian. Separating the cms motion and assuming zero total angular momentum ($S-$states), the Hamiltonian is reduced to one of six degrees of freedom in which the coordinates are the six $\rho-$variables. It is shown that the kinetic energy term is gauge-equivalent to a Laplace-Beltarmi operator plus an effective potential. In Section \ref{4bodyChain}, we focus on the 4-body closed chain of harmonic oscillators. It will be demonstrated that in the $\rho-$representation this system possesses a hidden $sl(7;R)$ Lie algebra structure. The relevant algebraic operator that governs the exact (polynomial) solutions of the problem is derived explicitly. Afterwards, in Section \ref{limitingcases} we study in detail the particular cases of four identical particles as well as the atomic-like and molecular-like (two-center and three-center) cases. The symmetry analysis (first and second order symmetries) of the free problem is presented in Section \ref{symmetryanalysis}. Finally, in Section \ref{reduction} we will address the degeneration to the plane $d=2$ and to the line $d=1$ where the number of independent $\rho-$variables reduces from 6 to 5 and 3, respectively. This makes the cases $d=1,2$ quite distinct from $d\geq 3$. For a certain class of potentials a further \emph{reduction} of the problem to one with less degrees of freedom is accomplished. The key idea is to introduce new modified, mass-dependent variables which are inspired in the geometric properties of the \emph{tetrahedron of interaction} whose vertices correspond to the positions of the particles. For conclusions and future outlook see Section \ref{conclusions}.

\section{Generalities}
\label{generalities}

The kinetic energy for the $4$-body quantum system of $d$-dimensional particles is of the form,
\begin{equation}
\label{Tflat}
   {\cal T}\ =\ -\,\bigg( \frac{1}{2\,m_1}\De_1^{(d)} \ + \ \frac{1}{2\,m_2}\De_2^{(d)} \ + \ \frac{1}{2\,m_3}\De_3^{(d)} \ + \ \frac{1}{2\,m_4}\De_4^{(d)}        \, \bigg) \ ,
\end{equation}
with coordinate vector of $i$th particle ${\bf r}_i \equiv {\bf r}^{(d)}_i=(x_{i,1}\,,\cdots \,,x_{i,d})$ and mass $m_i$. Here, $\De_i^{(d)}$ is the individual $d$-dimensional Laplacian,
\[
     \De_i^{(d)}\ =\ \frac{\pa^2}{\pa{{\bf r}_i} \pa{{\bf r}_i}}\ ,
\]
associated with the $i$th particle. The corresponding quantum Hamiltonian is given by,
\begin{equation}
\label{Hamiltonian}
  {\cal H}\ =\ {\cal T}\ +\ V\ .
\end{equation}
The configuration space for ${\cal H}$ is ${\mathbb{R}}^{4d}$. We assume that $V$ in (\ref{Hamiltonian}) is a scalar translational-invariant potential. Hence, the center-of-mass motion described by $d$-dimensional vectorial coordinate
\begin{equation}
\label{CMS}
    {\bf R}_{_0} \ =\ \frac{1}{\sqrt{M}}\,\sum_{k=1}^{4} m_k \,{\bf r}_{_k}\ , \qquad M\,=\,m_1\,+\,m_2\,+\,m_3\,+\,m_4\ ,
\end{equation}
can be separated out. After separation of the center-of-mass coordinate, the kinetic energy ${\cal T}$ in the space of relative motion ${\mathbb{R}}_r \equiv {\mathbb{R}}^{3 d }$ is described by the flat-space Laplacian $\De_r^{(3d)}$ (see below).

Let us introduce the three $d$-dimensional vectorial Jacobi coordinates

\begin{equation}
\label{Jacobi}
     {\bf r}^{(J)}_{j} \ = \ \sqrt{\frac{m_{j+1}M_j}{M_{j+1}}}\left({\bf r}_{j+1}\ - \ \sum_{k=1}^j\frac{m_k\,{\bf r}_k}{M_j}\right) \ ,
        \qquad \qquad \,j\,=\,1,2,3 \  ,
\end{equation}
($M_j = \sum_{k=1}^j m_k$) see e.g. \cite{Delves:1960} and also \cite{Willard:2018} for discussion. Remarkably, if the space of relative motion ${\mathbb{R}}_r$ is parameterized by these Jacobi coordinates the $3d$-dimensional Laplacian $\De_r^{(3d)}$ of the relative motion becomes diagonal,
\begin{equation}
\label{Tflat-diag}
   {\cal T}\ =\ -\sum_{i=1}^4\frac{1}{m_i}\De_i^{(d)} \ = \ -\frac{\pa^2}{\pa{{\bf R}_0} \pa{{\bf R}_0}}\  -  \
       \sum_{i=1}^{3}\frac{\pa^2}{\pa{{\bf r}_i^{(J)}} \pa{{\bf r}_i^{(J)}}} \ \equiv \ -\De_{{\bf R}_0} \ - \ \De_r^{(3d)} \ ,
\end{equation}
where $-\De_{{\bf R}_0}=-\frac{\pa^2}{\pa{{\bf R}_0} \pa{{\bf R}_0}}$ is the kinetic energy of the center-of-mass motion.
Note that the first Jacobi coordinate ${\bf r}^{(J)}_{1}$ is always proportional to the relative vector between particles 1 and 2.
Evidently, the Jacobi variables in ${\cal T}_r \equiv -\De_r^{(3d)}$ are separated and the kinetic energy of relative motion is the sum of kinetic energies in the Jacobi coordinate directions. By adding to ${\cal T}_r$ the harmonic
oscillator potentials in each Jacobi coordinate direction we arrive at the $4$-body Hamiltonian,
\begin{equation}
\label{H-osc-diag}
   {\cal H}_{r}\ =\   \
       \sum_{i=1}^{3} \bigg( \ -\frac{\pa^2}{\pa{{\bf r}_i^{(J)}} \pa{{\bf r}_i^{(J)}}}  + A_i\, \om^2\, { { ({\bf r}_i^{(J)}} } \cdot { {\bf r}_i^{(J)}})  \bigg)\ ,
\end{equation}
where $\om$ is the frequency and $A_i \geq 0,\ i=1,2,3$ play the role of spring constants. It describes 3 individual harmonic oscillators. Thus,
it is evident that this is an exactly-solvable problem: all eigenfunctions and eigenvalues are known analytically. We call this system the 4-body {\it Jacobi harmonic oscillator}. Let us note that if the potential in (\ref{H-osc-diag}) is chosen in the form of the moment of inertia $V=\sum^4_{i=1} m_i~{\bf r}_i^2$ all spring coefficients must be equal to each other and also to a reduced mass of the system, $A_i\, =\, \mu \, \equiv \, \bigg(\frac{m_1\,m_2\,m_3\,m_4}{M}\bigg)^{\frac{1}{3}}$, see e.g. \cite{Delves:1960,Fortunato:2017}. Thus, taking the moment of inertia as the potential $V$ in (\ref{Hamiltonian}) leads to the isotropic Jacobi harmonic oscillator (\ref{H-osc-diag}) with $A_i = \mu$ ($i=1,2,3$).
After the center-of-mass motion is removed,  the spectrum of (\ref{H-osc-diag}) is the sum of spectra of individual oscillators. Total zero angular momentum $L=0$ implies zero angular momenta of individual oscillators, hence the radial Hamiltonian of relative motion
is the sum of three $d$-dimensional radial Hamiltonians,
\begin{equation}
\label{H-osc-diag-radial}
   {\cal H}_{r}^{(L=0)}\ =\
       \sum_{i=1}^{3} \bigg(\ -\frac{\pa^2}{\pa{{r}_i^{(J)}} \pa{{r}_i^{(J)}}} \ - \
       \frac{(d-1)}{{r}_i^{(J)}}\,\frac{\pa}{\pa{{r}_i^{(J)}}}
       \ +\ A_i \,\om^2\, \big[\,r^{(J)}_i\,\big]^2\ \bigg)\ .
\end{equation}
Needless to say, the problem (\ref{H-osc-diag-radial}) is exactly-solvable (ES), its eigenfunctions are the product of individual eigenfunctions and the spectrum is linear in radial quantum numbers. Replacing the individual quadratic potential $A_i \,\om^2\, (r^{(J)}_i)^2$ by the  quasi-exactly-solvable (QES) sextic potential, we arrive at the QES anharmonic Jacobi oscillator.
It is easy to check that in the $3$-dimensional space of relative radial motion of modules squared of Jacobi coordinates, or, saying differently, of the Jacobi distances squared ${\rho}^{(J)}_{j}=|{\bf r}^{(J)}_{j}|^2$,  its hidden algebra is $sl_2^{\,\otimes \,{3}}$  acting on this  $3$-dimensional space.

Since the spectrum of the Jacobi oscillators (\ref{H-osc-diag}), (\ref{H-osc-diag-radial}) is known explicitly, their eigenfunctions can be used as the basis to study many-body problems, as was proposed in \cite{Delves:1960}.
The present authors are not aware of any studies of the Jacobi oscillators {\it per se}.

In this work we will explore the case of a 4-body system with quadratic (harmonic) and sextic potentials which depend on relative {\it distances squared}, ${|{\bf r}_i - {\bf r}_j |}^2$, c.f. (\ref{Jacobi}), between particles only.

\section{Four body system}
\label{4bodyS}

The general quantum Hamiltonian for four $d$-dimensional ($d>2$) bodies of masses $m_1, m_2, m_3, m_4$
with potential that solely depends on relative (mutual) distances
between particles, is of the form,
\begin{equation}
\label{Hgen}
   {\cal H}\ =\ -\sum_{i=1}^4 \frac{1}{2 \,m_i} \De_i^{(d)}\ +\  V(r_{12},\,r_{13},\,r_{14},\,r_{23},\,r_{24},\,r_{34})\ ,\
\end{equation}
see e.g. \cite{Willard:2018,MTE:2018}, where $\De_i^{(d)}$ is $d$-dimensional Laplacian of $i$th particle with coordinate vector ${\bf r}_i \equiv {\bf r}^{(d)}_i=(x_{i,1},\, x_{i,2},\,x_{i,3},\ldots,\,x_{i,d})$\ , and
\begin{equation}
\label{rel-coord}
r_{ij}\ = \ |{\bf r}_i\, -\, {\bf r}_j|\ ,\qquad i,j=1,2,3,4\ ,
\end{equation}
is the (relative) distance between particles $i$ and $j$, $r_{ij}=r_{ji}$. Separating the center-of-mass (\ref{CMS}) and then making the change of variables in the space of relative motion ${\mathbb{R}}_r= {\mathbb{R}}^{3 d }$ from Jacobi-coordinates (\ref{Jacobi}) to generalized Euler coordinates (six relative distances (\ref{rel-coord}) \emph{squared} and $(3d-6)$ \emph{angles} ${\Om}$)
\[
   (\,{\bf r}^{(J)}_{1}\ ,\ {\bf r}^{(J)}_{2}\ ,\ {\bf r}^{(J)}_{3}\,)\ \rar \ (\,\rho_{12}\ ,\ \rho_{13}\ , \ \rho_{14}\ ,\ \rho_{23}\ ,\ \rho_{24}\ ,\ \rho_{34}\ ,   \ {\Om}\,)\ ,
\]
with
\[
\rho_{ij} \ \equiv \  r_{ij}^2  \ = \ {|{\bf r}_i\, -\, {\bf r}_j|}^2     \ ,\qquad i,j=1,2,3,4\ ,
\]
we arrive at a six-dimensional radial-type Schr\"odinger equation \cite{MTE:2018}
\begin{equation}
\label{H-3-body}
       \big[\,-\Delta_{\rm rad}(\rho)\ + \ V(\rho)\,\big]\,\Psi(\rho)\ =\ E \,\Psi(\rho)\ ,\
\end{equation}
where
{\small
\begin{equation}
\begin{aligned}
\label{addition3-3r-M}
& \Delta_{\rm rad}(\rho_{ij}, \pa_{ij})\ =  \   2\bigg( \frac{1}{\mu_{12}} \rho_{12} \,\pa^2_{\rho_{12}} + \frac{1}{\mu_{13}}\rho_{13}\, \pa^2_{\rho_{13}} +\frac{1}{\mu_{14}}\rho_{14}\, \pa^2_{\rho_{14}}
 +\frac{1}{\mu_{23}}\rho_{23}\, \pa^2_{\rho_{23}} +\frac{1}{\mu_{24}}\rho_{24}\, \pa^2_{\rho_{24}}
 \\ &  +\frac{1}{\mu_{34}}\rho_{34}\, \pa^2_{\rho_{34}} \bigg) + d\,\bigg(\frac{1}{\mu_{12}}\pa_{\rho_{12}} + \frac{1}{\mu_{13}}\pa_{\rho_{13}}+ \frac{1}{\mu_{14}}\pa_{\rho_{14}}+ \frac{1}{\mu_{23}}\pa_{\rho_{23}}+ \frac{1}{\mu_{24}}\pa_{\rho_{24}}+ \frac{1}{\mu_{34}}\pa_{\rho_{34}}\bigg)
\\ &
 +  \frac{2}{m_1} \bigg({(\rho_{12} + \rho_{13} - \rho_{23})}\pa_{\rho_{12}}\pa_{\rho_{13}}\ +
          {(\rho_{12} + \rho_{14} - \rho_{24})}\pa_{\rho_{12}}\pa_{\rho_{14}}\ +
          {(\rho_{13} + \rho_{14} - \rho_{34})}\pa_{\rho_{13}}\pa_{\rho_{14}} \bigg)
\\ &
+  \frac{2}{m_2} \bigg((\rho_{12} + \rho_{23} - \rho_{13})\pa_{\rho_{12}}\pa_{\rho_{23}}\ +
          (\rho_{12} + \rho_{24} - \rho_{14})\pa_{\rho_{12}}\pa_{\rho_{24}}\ +
          (\rho_{23} + \rho_{24} - \rho_{34})\pa_{\rho_{23}}\pa_{\rho_{24}}
    \bigg)
\\ &
+  \frac{2}{m_3} \bigg((\rho_{13} + \rho_{23} - \rho_{12})\pa_{\rho_{13}}\pa_{\rho_{23}}\ +
          (\rho_{13} + \rho_{34} - \rho_{14})\pa_{\rho_{13}}\pa_{\rho_{34}}\ +
          (\rho_{23} + \rho_{34} - \rho_{24})\pa_{\rho_{23}}\pa_{\rho_{34}}
    \bigg)
\\ &
+  \frac{2}{m_4} \bigg((\rho_{14} + \rho_{24} - \rho_{12})\pa_{\rho_{14}}\pa_{\rho_{24}}\ +
          (\rho_{14} + \rho_{34} - \rho_{13})\pa_{\rho_{14}}\pa_{\rho_{34}}\ +
          (\rho_{24} + \rho_{34} - \rho_{23})\pa_{\rho_{24}}\pa_{\rho_{34}}
    \bigg)      \ ,
\end{aligned}
\end{equation}
}
and
\[
   \frac{1}{\mu_{ij}}\ = \ \frac{m_i+m_j}{m_i\, m_j}\ ,
\]
is the inverse reduced mass, which governs six-dimensional (radial) dynamics in $\rho-$variables. We call it $\rho-$representation. In these variables, the operator (\ref{addition3-3r-M}) is not $S_6$ permutationally-invariant. Nevertheless, it remains $S_4$ invariant under the permutations of the particles.

The six-dimensional operator in (\ref{H-3-body})
\begin{equation}
\label{Hrad3}
  H_{\rm rad}\ \equiv\ -\De_{\rm rad} \ + \  V\ ,
\end{equation}
is, in fact, gauge-equivalent to a Schr\"odinger operator, see \cite{MTE:2018}. It can be called six-dimensional radial Hamiltonian. In particular, the operator $\Delta_{\rm rad}(\rho)$ is self-adjoint with respect to the normalized measure given by

\begin{equation}\label{}
dv_{\rm rad} \ = \ V_4^{d-4}\,d\rho_{12}\,d\rho_{13}\,d\rho_{14}\,d\rho_{23}\,d\rho_{24}\,d\rho_{34}\ .
\end{equation}

\begin{figure}[htp]
  \centering
  \includegraphics[width=14.0cm]{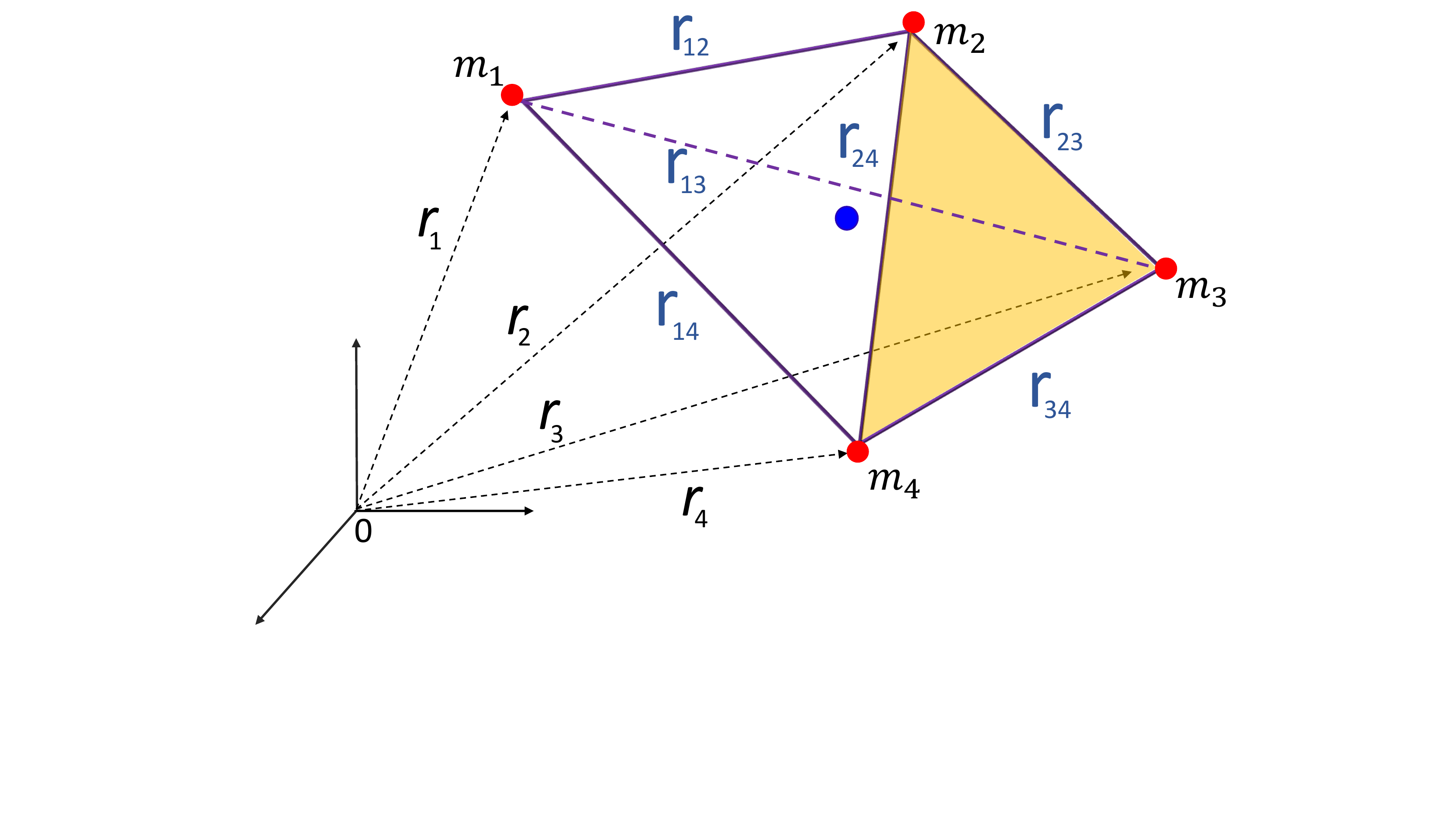}
  \caption{Tetrahedron of interaction in $3D$ space. Center-of-mass (the barycenter of tetrahedron) is marked by (blue) bubble. Each of the four faces forms a triangle of interaction.}
\label{Fig1}
\end{figure}

The corresponding configuration space (the physics domain) is confined to the hypercube ${\bf R}_+(\rho_{12}) \times {\bf R}_+(\rho_{13}) \times {\bf R}_+(\rho_{14})\times {\bf R}_+(\rho_{23}) \times {\bf R}_+(\rho_{24}) \times {\bf R}_+(\rho_{34})$ in $E_6$. More explicitly, it is given by the conditions
\be
\label{PhyDom}
0 < \rho_{A},\rho_{B},\rho_{C} < \infty\ ,\quad
{\rho}_{A} <  (\sqrt{{\rho}_{B}} + \sqrt{{\rho}_{C}})^2,\quad {\rho}_{B} < (\sqrt{{\rho}_{A}} + \sqrt{{\rho}_{C}})^2,\quad {\rho}_{C} < \ (\sqrt{{\rho}_{A}} + \sqrt{{\rho}_{B}})^2 \ ,
\ee
($ A \neq B \neq C = 12,13,14,23,24,34 $). We remark that
\begin{equation}
\label{CFrho}
\quad
S^2_{\Delta ABC} \equiv  \frac{ 2 \,(\rho _{A}\, \rho _{B}+  \rho _{A}\, \rho _{C}+
       \rho _{B}\, \rho _{C})  - (\rho _{A}^2+\rho _{B}^2+\rho _{C}^2) }{16} \ \geq \ 0   \ ,
\end{equation}
because the left-hand side (l.h.s.) written in terms of relative distances between particles is equal to
\be\frac{1}{16}(r_{A}+r_{B}-r_{C})(r_{A}+r_{C}-r_{B})(r_{B}+r_{C}-r_{A})(r_{A}+r_{B}+r_{C})\ ,\ee
and conditions (\ref{PhyDom})  hold. Therefore, from the Heron formula, $S^2_{\Delta ABC}$ is the square of the area of the \emph{triangle of interaction} with sides $r_{A},\,r_B$ and $r_C$ \,. The triangles of interaction are nothing but the faces of the \emph{tetrahedron of interaction}, see Fig.\ref{Fig1}. The conditions (\ref{PhyDom}) state that the square of the area of the tetrahedron formed by the particle positions must be positive.

\subsection{Laplace-Beltrami operator: underlying geometry}

The remarkable property of the algebraic operator ${\De_{\rm rad}}(\rho)$ (\ref{addition3-3r-M}) is its gauge-equivalence to the Schr\"odinger operator (see below) with the co-metric
{\small
\begin{equation}
\label{gmn33-rho}
 g^{\mu \nu}(\rho)\ = \left(
\begin{array}{cccccc}
 \frac{2}{\mu_{12}}\, \rho _{12} & \frac{\rho _{12}+\rho _{13}-\rho _{23}}{m_1} & \frac{\rho _{12}+\rho _{14}-\rho _{24}}{m_1} & \frac{\rho _{12}-\rho _{13}+\rho _{23}}{m_2} & \frac{\rho _{12}-\rho _{14}+\rho _{24}}{m_2} & 0 \\
 \frac{\rho _{12}+\rho _{13}-\rho _{23}}{m_1} & \frac{2}{\mu_{13}}\, \rho _{13} & \frac{\rho _{13}+\rho _{14}-\rho _{34}}{m_1} & \frac{\rho _{13}+\rho _{23}-\rho _{12}}{m_3} & 0 & \frac{\rho _{13}-\rho _{14}+\rho _{34}}{m_3} \\
 \frac{\rho _{12}+\rho _{14}-\rho _{24}}{m_1} & \frac{\rho _{13}+\rho _{14}-\rho _{34}}{m_1} & \frac{2}{\mu_{14}} \,\rho _{14} & 0 & \frac{\rho _{14}+\rho _{24}-\rho _{12}}{m_4} & \frac{\rho _{14}+\rho _{34}-\rho _{13}}{m_4} \\
 \frac{\rho _{12}-\rho _{13}+\rho _{23}}{m_2} & \frac{\rho _{13}+\rho _{23}-\rho _{12}}{m_3} & 0 & \frac{2}{\mu_{23}} \,\rho _{23} & \frac{\rho _{23}+\rho _{24}-\rho _{34}}{m_2} & \frac{\rho _{23}-\rho _{24}+\rho _{34}}{m_3} \\
 \frac{\rho _{12}-\rho _{14}+\rho _{24}}{m_2} & 0 & \frac{\rho _{14}+\rho _{24}-\rho _{12}}{m_4} & \frac{\rho _{23}+\rho _{24}-\rho _{34}}{m_2} & \frac{2}{\mu_{24}}\, \rho _{24} & \frac{\rho _{24}+\rho _{34}-\rho _{23}}{m_4} \\
 0 & \frac{\rho _{13}-\rho _{14}+\rho _{34}}{m_3} & \frac{\rho _{14}+\rho _{34}-\rho _{13}}{m_4} & \frac{\rho _{23}-\rho _{24}+\rho _{34}}{m_3} & \frac{\rho _{24}+\rho _{34}-\rho _{23}}{m_4} & \frac{2}{\mu_{34}}\, \rho _{34} \\
\end{array}
\right) \ .
\end{equation}
}

In (\ref{gmn33-rho}), we have made the identifications $1 \rightarrow \rho _{12},\,2 \rightarrow \rho _{13},\,3 \rightarrow \rho _{14},\,4 \rightarrow \rho _{23},\,5 \rightarrow \rho _{24},\,6 \rightarrow \rho _{34},\,$ for $\mu$ and $\nu$. Its determinant

\begin{equation}
\label{gmn33-rho-det-M}
D_m\ =\ \det g^{\mu \nu}\ =\ 9216\,c_m\,V_4^2\,\bigg[ \big(\sum V_{2,m}\big)\big( \sum V_{3,m}\big)\, -\, 9\,(m_1+m_2+m_3+m_4)\,V_4^2      \bigg]  \ ,
\end{equation}
is positive definite, where $c_m=\frac{m_1+m_2+m_3+m_4}{m_1^2\,m_2^2\,m_3^2\,m_4^2}$,
\begin{equation}
\label{V4}
\begin{aligned}
& V_4^2  \  = \  \frac{1}{144} \bigg[\,\left[\left(\rho _{13}+\rho _{14}+\rho _{23}+
  \rho _{24}\right) \rho _{34}-\left(\rho _{13}-\rho _{14}\right) \left(\rho _{23}-\rho _{24}\right)-
  \rho _{34}^2 \right] \rho _{12}
\\ & - \ \rho _{13}^2 \rho _{24} \ - \ \rho _{34} \rho _{12}^2 \ +
\  \rho _{23} \left[\left(\rho _{14}-\rho _{24}\right) \rho _{34}-\rho _{14} \left(\rho _{14}+
  \rho _{23}-\rho _{24}\right)\right]
\\ &  \ + \ \rho _{13} \left[\,\rho _{14} \left(\rho _{23}+\rho _{24}-
  \rho _{34}\right)+\rho _{24} \left(\rho _{23}-\rho _{24}+\rho _{34}\right)\right]\bigg]  \ ,
\end{aligned}
\end{equation}
is the square of the volume of the \emph{tetrahedron of interaction},
\be
\label{Pweighted}
\sum V_{2,m}\ =\ m_1 m_2 \,\rho_{12}  \,+\,m_1 m_3 \,\rho_{13} \,+\,m_1 m_4 \,\rho_{14} \,+\,m_2 m_3 \,\rho_{23}\, +\,m_2 m_4 \,\rho_{24} \,+\,m_3 m_4 \,\rho_{34}  \ ,
\ee
is the weighted sum of square of sides and diagonals of the tetrahedron of interaction,
{\small
\be
\label{Sweighted}
\sum V_{3,m} \ = \  \frac{1}{m_1} S^2(\sqrt{{\rho_{23}}},\,\sqrt{{\rho_{24}}},\,\sqrt{{\rho_{34}}}) \ + \ \frac{1}{m_2} S^2(\sqrt{{\rho_{13}}},\,\sqrt{{\rho_{14}}},\,\sqrt{{\rho_{34}}})\ + \  \frac{1}{m_3} S^2(\sqrt{{\rho_{12}}},\,\sqrt{{\rho_{14}}},\,\sqrt{{\rho_{24}}})
\ee
\[
\ + \ \frac{1}{m_4} S^2(\sqrt{{\rho_{12}}},\,\sqrt{{\rho_{13}}},\,\sqrt{{\rho_{23}}}) \ ,
\]
}
is the weighted sum of squares of areas, here $ S^2(a,\,b,\,c)$ is the square of the area of the triangle of interaction with sizes $a,b,c$. Hence, $D_m$ is still proportional to $V_{4}^2$ (square of the volume of tetrahedron) being of a pure geometrical nature!

Making the gauge transformation of (\ref{addition3-3r-M}) with determinant (\ref{gmn33-rho-det-M}) as the factor,
\be
\label{GauFa}
         \Gamma \ = \  D_m^{-\frac{1}{4}}\,V_4^{1 - \frac{d}{4}} \quad ,
\ee
we find that
\begin{equation}
         \Gamma^{-1}\, {\De_{\rm rad}}(\rho)\,\Gamma \ =
        \  \De_{LB}(\rho) \ - \ V_{\rm eff}(\rho) \ ,
\label{HLB3Ma}
\end{equation}
is the Laplace-Beltrami operator with metric $g^{\mu \nu}$ (\ref{gmn33-rho})
\[
   \De_{LB}(\rho)\ = \ {\sqrt{D_m}}\ \pa_{\mu}\ \frac{1}{\sqrt{D_m}}\  g^{\mu \nu} \pa_{\nu}\ ,
\]
plus an effective potential given by
\be
\label{Veffgen}
V_{\rm eff}\ =\  \ \frac{3\,(\sum {V}_{2,m}^2)^2+28\,(m_1+m_2+m_3+m_4)\,m_1\,m_2\,m_3\,m_4\,\sum{ V}_{3,m}^2}{32\,m_1\,m_2\,m_3\,m_4\,((\sum { V}_{2,m}^2)\sum { V}_{3,m}^2-9\,(m_1+m_2+m_3+m_4)\,{ V}_4^2)}
\ee
\begin{equation}
+ \ \frac{(d-5)(d-3)\sum{V}_{3,m}^2}{72\,{V}_4^2}   \ ,
\nonumber
\end{equation}
where its 2nd term is absent for $d=3,5$. Thus, from the original Hamiltonian (\ref{Hgen}) we arrive at the spectral problem for the Schr\"odinger operator
\begin{equation}\label{HLB4b}
  H_{LB} \ = \  -\De_{LB} \ + \ V_{\rm eff} \ + \ V \ ,
\end{equation}
in the $\rho$-space with $d > 2$. The $d$-independent Laplace-Beltrami operator $-\De_{LB}$ plays the role of a kinetic energy. The Hamiltonian $H_{LB}$ describes a 6-dimensional quantum particle moving in curved space where $V_{\rm eff}$ can be considered as the centrifugal potential. The connection between the kinetic energy (${\cal T}=\sum_{i=1}^4 \frac{1}{2 \,m_i} \De_i^{(d)}$) in the original Hamiltonian (\ref{Hgen}) and that of the Hamiltonian $H_{LB}$ (\ref{HLB4b}) can be summarized as follows,
\[
{\cal T}\quad {}_{\overrightarrow{\text {removal of} \ {\bf R}_{0}} } \quad \Delta^{(3d)}_{r} \quad {}_{\overrightarrow{\text { angle-independent solutions} }} \quad \Delta^{}_{\rm rad} \quad {}_{\overrightarrow{\text {gauge transformation}\ \Gamma}} \quad \Delta^{}_{LB} \ .
\]
Consequently, we reduce the original $4d$-dimensional problem to a 6 dimensional one. As for the potential, we simply add the effective potential $V_{\rm eff}$ (\ref{Veffgen}) that arises from the $d-$dependent gauge transformation $\Gamma$ (\ref{GauFa}).

\subsection*{Classical Mechanics}

Making the \emph{de-quantization} of (\ref{HLB4b}) we arrive at a 6-dimensional classical system which is characterized by the classical Hamiltonian \footnote{Restoring $\hbar$, it turns out that $V_{\rm eff}$ (\ref{Veffgen}) is proportional to $\hbar^2$, thus, it vanishes in the classical limit},
\begin{equation}
\label{H-4-3r-rho-class}
    {\cal H}_{LB}^{(c)} (\rho) \ =\ g^{\mu \nu}(\rho)\,P_{\mu}\, P_{\nu} \  + \  V(\rho)  \ ,
\end{equation}
$\nu,\,\mu\,=\,12,13,14,23,24,34$, where $(\rho_{\mu},\,P_{\mu})$ are pairs of canonical conjugate variables whilst $g^{\mu \nu}(\rho)$ is given by (\ref{gmn33-rho}). The reduced Hamiltonian (\ref{H-4-3r-rho-class}) may be suitable for investigating special configurations (trajectories) or the relative equilibria of the classical 4-body problem \cite{Erdi,Hampton}.

\section{Four-body chain of harmonic oscillators}
\label{4bodyChain}

In this Section we focus on the four-body oscillator system at zero total angular momentum ($S$-states). This system is described by the Hamiltonian $H_{\rm rad}$ (\ref{Hrad3}) with potential
\begin{equation}
\label{V3-es}
   V^{(es)}\ =\ 2\,\om^2\bigg[   \nu_{12}\,\rho_{12} \ + \  \nu_{13}\,\rho_{13} \ + \  \nu_{14}\,\rho_{14} \ + \ \nu_{23}\,\rho_{23} \ + \  \nu_{24}\,\rho_{24} \ + \  \nu_{34}\,\rho_{34} \bigg]\ ,
\end{equation}
where $\om > 0$ and $\nu_{ij}$ are positive parameters that play the role of spring constants, see below. In this case, the equation (\ref{H-3-body}) possesses infinitely-many eigenfunctions. It is easy to check that the ground state function is given by
\begin{equation}
\label{Psi03}
   \Psi_0^{(es)}\ =\ e^{-\om\, (a\,\mu_{12}\,\rho_{12}\,+\,b\,\mu_{13}\,\rho_{13}\,+\,c\,\mu_{14}\,\rho_{14}\,+\,e\,\mu_{23}\,\rho_{23}\,+\,f\,\mu_{24}\,\rho_{24}
   \,+\,g\,\mu_{34}\,\rho_{34})}\ ,
\end{equation}
where $a,\,b,\,c,\,e,\,f,\,g$ are determined by the spring constants $\nu_{ij}$ via the nonlinear algebraic relations
\begin{equation}
\label{freq-3}
\begin{aligned}
&  \nu_{12} \ = \  a^2\, \mu _{12} \ + \ a\, b\ \frac{\mu _{12}\, \mu _{13} }{m_1} \ + \  a\, c\ \frac{\mu _{12} \,\mu _{14} }{m_1}\ + \  a\, e\ \frac{\mu _{12} \,\mu _{23} }{m_2}\ + \  a\, f\ \frac{\mu _{12} \,\mu _{24} }{m_2}
\\ &
\ - \  b\, e \ \frac{\mu_{13} \,\mu _{23} }{m_3}\ - \  c\, f \ \frac{\mu_{14} \,\mu _{24} }{m_4} \ ,
\\ &
\nu_{13} \ = \  b^2\, \mu _{13} \ + \ b\, a\ \frac{\mu _{13}\, \mu _{12} }{m_1} \ + \  b\, c\ \frac{\mu _{13} \,\mu _{14} }{m_1}\ + \  b\, e\ \frac{\mu _{13} \,\mu _{23} }{m_3}\ + \  b\, g\ \frac{\mu _{13} \,\mu _{34} }{m_3}
\\ &
\ - \  a\, e \ \frac{\mu_{12} \,\mu _{23} }{m_2}\ - \  c\, g \ \frac{\mu_{14} \,\mu _{34} }{m_4} \ ,
\\ &
\vdots
\\ &
\nu_{34} \ = \  g^2\, \mu _{34} \ + \ g\, b\ \frac{\mu _{34}\, \mu _{13} }{m_3} \ + \  g\, c\ \frac{\mu _{34} \,\mu _{14} }{m_4}\ + \  g\, e\ \frac{\mu _{34} \,\mu _{23} }{m_3}\ + \  g\, f\ \frac{\mu _{34} \,\mu _{24} }{m_4}
\\ &
\ - \  b\, c \ \frac{\mu_{13} \,\mu _{14} }{m_1}\ - \  e\, f \ \frac{\mu_{23} \,\mu _{24} }{m_4}
\ .
\end{aligned}
\end{equation}
Normalizability of $\Psi_0^{(es)}$ requires all roots $a,\,b,\,c,\,e,\,f,\,g$ to be positive. The ground state energy
\begin{equation}
\label{E0en}
E_0\ = \ \om \,d\,(a+b+c+e+f+g) \ ,
\end{equation}
is proportional to the dimension $d$.

\begin{figure}[htp]
  \centering
  \includegraphics[width=8.0cm]{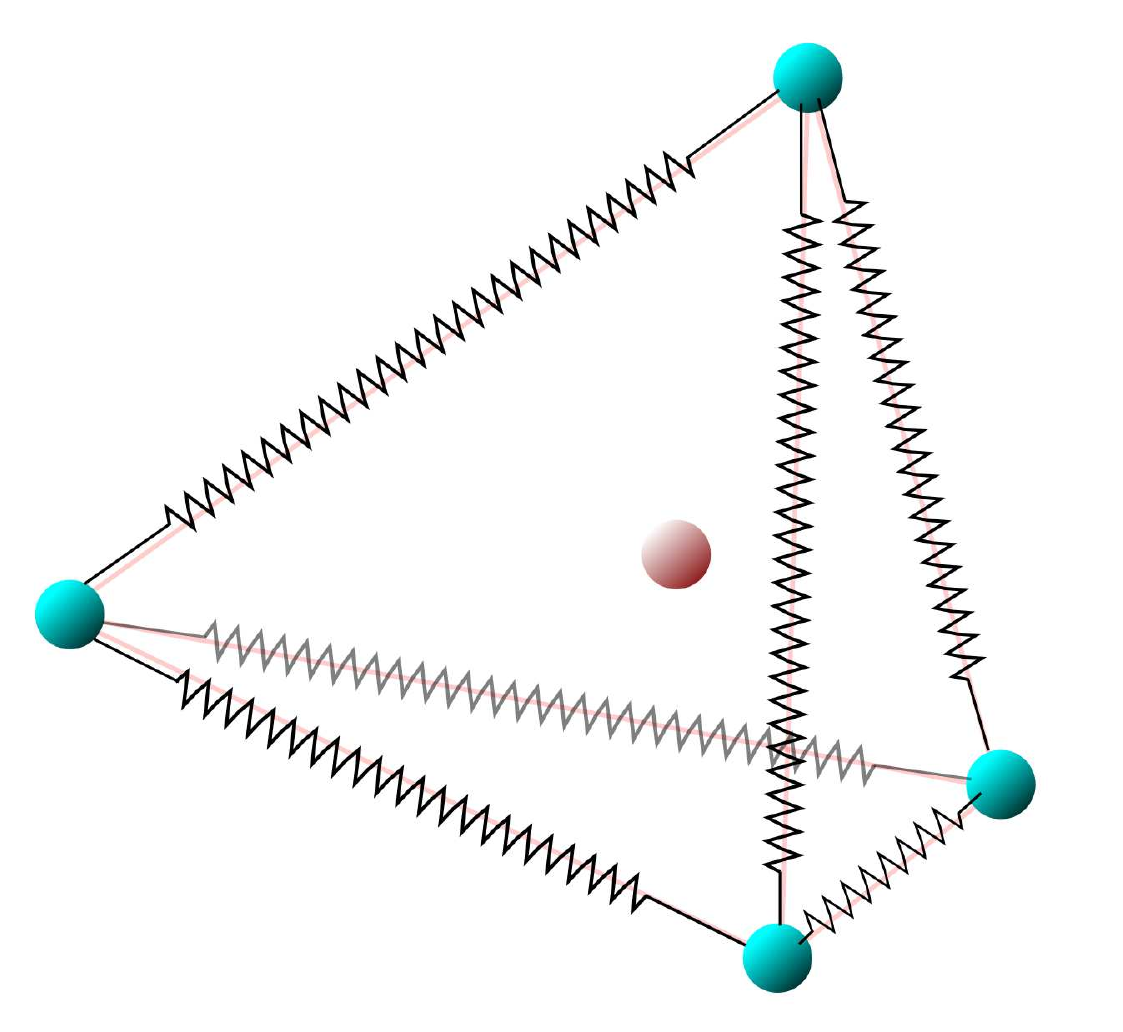}
  \caption{Four-body harmonic oscillator chain. The harmonic potential $V^{(es)}$ (\ref{V3-es})
admits infinitely-many eigenfunctions.   }
  \label{3-body}
\end{figure}

\clearpage

By making a gauge rotation with $\Psi_0^{(es)}$ (\ref{Psi03}) as the factor, we convert $H_{\rm rad}$ (\ref{Hrad3}) to an algebraic differential operator where the coefficient functions are polynomials. Explicitly,
\begin{equation}\label{H-4body-algebraic}
h^{(es)} \equiv \big( \Psi_0^{(es)}\big)^{-1}\, [-\De_{\rm rad}(\rho) + V^{(es)}(\rho) - E_0]\, \Psi_0^{(es)}
 \ = \ -\De_{\rm rad}(\rho) \ + \ 2\,\omega\,\nabla_{\rm rad}
\end{equation}
here the first-order differential operator
{\small
\begin{equation}
\begin{aligned}
&\nabla_{\rm rad} \ = \
\bigg[ \,2\,a\,\rho_{12} \ + \   \frac{\mu_{13}\,b\, \left(\rho _{12}+\rho _{13}-\rho _{23}\right)}{m_1}\ + \ \frac{\mu_{23}\,e \left(\rho _{12}-\rho _{13}+\rho _{23}\right)}{m_2} \ + \
\\ &
\frac{\mu_{14}\,c \,\left(\rho _{12}+\rho _{14}-\rho _{24}\right)}{m_1}+\frac{\mu_{24}\,f \left(\rho _{12}-\rho _{14}+\rho _{24}\right)}{m_2}    \bigg]  \,\pa_{\rho_{12}}  \ + \ \bigg[ \,2\,b\,\rho_{13}  \ + \
\\ &
\ + \ \frac{\mu_{12}\,a\, \left(\rho _{12}+\rho _{13}-\rho _{23}\right)}{m_1}\ + \ \frac{\mu_{14}\,c \left(\rho _{13}-\rho _{34}+\rho _{14}\right)}{m_1} \ + \   \frac{\mu_{23}\,e\, \left(\rho _{23}+\rho _{13}-\rho _{12}\right)}{m_3} \ + \
\\ &
 \frac{\mu_{34}\,g \left(\rho _{34}-\rho _{14}+\rho _{13}\right)}{m_3}  \bigg]  \,\pa_{\rho_{13}} \ + \ \bigg[ \,2\,c\,\rho_{14}  \ + \  \frac{\mu_{12}\,a \left(\rho _{14}+\rho _{12}-\rho _{24}\right)}{m_1}  \ + \
\\ &
 \frac{\mu_{13}\,b \left(\rho _{14}+\rho _{13}-\rho _{34}\right)}{m_1} \ +  \ \frac{\mu_{24}\,f \left(\rho _{14}+\rho _{24}-\rho _{12}\right)}{m_4} \ + \ \frac{\mu_{34}\,g \left(\rho _{14}+\rho _{34}-\rho _{13}\right)}{m_4}    \bigg]  \,\pa_{\rho_{14}} \  + \
\\ &
\bigg[ \,2\,e\,\rho_{23} \ + \  \frac{\mu_{12}\,a \left(\rho_{23}+\rho _{12}-\rho _{13}\right)}{m_2}    \ + \ \frac{\mu_{13}\,b \left(\rho_{23}+\rho _{13}-\rho _{12}\right)}{m_3}    \ + \
\\ &
  \frac{\mu_{24}\,f \left(\rho_{23}+\rho _{24}-\rho _{34}\right)}{m_2}    \ + \  \frac{\mu_{34}\,g \left(\rho_{23}+\rho _{34}-\rho _{24}\right)}{m_3}   \bigg]  \,\pa_{\rho_{23}}   \ + \ \bigg[ \,2\,f\,\rho_{24}  \ + \
\\ &
\ + \  \frac{\mu_{12}\,a \left(\rho_{24}+\rho _{12}-\rho _{14}\right)}{m_2}  \ + \ \frac{\mu_{14}\,c \left(\rho_{24}+\rho _{14}-\rho _{12}\right)}{m_4}  \ + \ \frac{\mu_{23}\,e \left(\rho_{24}+\rho _{23}-\rho _{34}\right)}{m_2}  \ + \
\\ &
  \frac{\mu_{34}\,g \left(\rho_{24}+\rho _{34}-\rho _{23}\right)}{m_4} \bigg]  \,\pa_{\rho_{24}}  \ + \ \bigg[ \,2\,g\,\rho_{34}   \ + \  \frac{\mu_{13}\,b \left(\rho _{34}+\rho _{13}-\rho _{14}\right)}{m_3}  \ + \
\\ &
 \frac{\mu_{14}\,c \left(\rho _{34}+\rho _{14}-\rho _{13}\right)}{m_4}  \ +  \ \frac{\mu_{23}\,e \left(\rho _{34}+\rho _{23}-\rho _{24}\right)}{m_3}  \ + \  \frac{\mu_{24}\,f \left(\rho _{34}+\rho _{24}-\rho _{23}\right)}{m_4}    \bigg]  \,\pa_{\rho_{34}} \ ,
\end{aligned}
\end{equation}

and $E_0$ is given by (\ref{E0en}). The algebraic operator $h^{(es)}$ (\ref{H-4body-algebraic}) preserves the space of
polynomials
\begin{equation}
\label{P6}
     {\mathcal P}^{(1,1,1,1,1,1)}_{N}\ =\ \langle \rho_{12}^{p_1}\ \rho_{13}^{p_2} \ \rho_{14}^{p_3} \ \rho_{23}^{p_4}\ \rho_{24}^{p_5} \ \rho_{34}^{p_6} \vert \
     0 \le p_1 + p_2+ p_3 +p_4+p_5+p_6 \le N \rangle\ ,
\end{equation}
for any positive integer $N=0,1,2, \ldots $. Hence, it preserves the flag ${\mathcal P}_N^{(1,1,1,1,1,1)}$ with the characteristic (weight) vector $(1,1,1,1,1,1)$.

\clearpage

\subsection{Hidden $sl(7,{\bf R})$ Lie algebraic structure}

Note that the operator (\ref{H-4body-algebraic}) is of Lie-algebraic nature: it can be rewritten in terms of the generators  of the maximal affine sub-algebra  $b_7$ of the algebra $sl(7,{\bf R})$ realized by the following first order differential operators, see e.g. \cite{Turbiner:1988}
\begin{eqnarray}
\label{sl4R}
 {\cal J}_i^- \ &=& \ \frac{\pa}{\pa u_i}\ ,\qquad \quad i=1,2,\ldots, 6\ , \non  \\
 {{\cal J}^0_{ij}} \ &=& \
               u_i \frac{\pa}{\pa u_j}\ , \qquad i,j=1,2,\ldots, 6 \ ,\non \\
 {\cal J}^0(N)\  &=& \ \sum_{i=1}^{6} u_i \frac{\pa}{\pa u_i}-N\, , \non \\
 {\cal J}_i^+(N) \ &=& \ u_i \,{\cal J}^0(N)\ = \
    u_i\, \left( \sum_{j=1}^{6} u_j\frac{\pa}{\pa u_j}-N \right)\ ,
       \quad i=1,2,\ldots, 6 \ ,
\end{eqnarray}
where $N$ is an arbitrary parameter and we have introduced the notation
\[
 u_1\equiv\rho_{12}\ ,\qquad u_2\equiv\rho_{13}\ , \qquad u_3\equiv\rho_{14} \ , \qquad u_4\equiv\rho_{23}\ ,\qquad u_5\equiv\rho_{24}\ , \qquad u_6\equiv\rho_{34} \ .
\]
If $N$ is a non-negative integer, a finite-dimensional representation space appears in the form
\begin{equation}
\label{P3-sl4}
     {\cal P}_{N}\ =\ \langle u_1^{p_1}\, u_2^{p_2}\, u_3^{p_3}\,u_4^{p_4}\, u_5^{p_5}\, u_6^{p_6} \vert \ 0 \le p_1+p_2+p_3+p_4+p_5+p_6 \le N \rangle\ .
\end{equation}
This space coincides with (\ref{P6}). It is easy to check that the space ${\cal P}_{N}$ is invariant with respect to $6D$ projective transformations,
\[
   u_i \rar \frac{\sum_{j=1}^{6}\alpha_{ij}\,u_j+ \gamma_{i}}
                 {\sum_{j=1}^{6}\beta_j\,u_j + \kappa }\qquad ,\qquad i\,=\,1,2,\ldots,6 \ ,
\]
where $\alpha_{ij}, \beta_{j}, \gamma_{i}, \kappa$ are real parameters. By taking the parameters $\alpha_{ij},\gamma_{i}$ at $i=1,2,\ldots,6$  and $\beta_j, \kappa$ as the rows of the $7 \times 7$ matrix $G$ one can demonstrate that $G \in GL(7)$.

The spectrum of (\ref{H-4body-algebraic}) depends on six integers (quantum numbers) and is linear in quantum numbers. In terms of the generators (\ref{sl4R}), the algebraic operator (\ref{H-4body-algebraic}) takes the form

\begin{equation}
\label{H4LieA}
\begin{aligned}
 h^{(es)}({\cal J}) & \ = \ - 2\bigg[\, \frac{1}{\mu_{12}}{\cal J}_{11}^0\,{\cal J}_1^- \ + \  \frac{1}{\mu_{13}}{\cal J}_{22}^0\,{\cal J}_2^-
     \ + \  \frac{1}{\mu_{14}}{\cal J}_{33}^0\,{\cal J}_3^- \ + \    \frac{1}{\mu_{23}}{\cal J}_{44}^0\,{\cal J}_4^- \ + \ \frac{1}{\mu_{24}}{\cal J}_{55}^0\,{\cal J}_5^- \ + \
\\ &
   \frac{1}{\mu_{34}}{\cal J}_{66}^0\,{\cal J}_6^-  \,\bigg]  \ - \
d \,\bigg[\,\frac{1}{\mu_{12}}{\cal J}_1^- + \frac{1}{\mu_{13}}{\cal J}_2^- + \frac{1}{\mu_{14}}{\cal J}_3^- + \frac{1}{\mu_{23}}{\cal J}_4^-  + \frac{1}{\mu_{24}}{\cal J}_5^-  + \frac{1}{\mu_{34}}{\cal J}_6^- \,\bigg] \ - \
\\ &
   \frac{2}{m_1} \bigg[{({\cal J}_{11}^0 + {\cal J}_{21}^0 - {\cal J}_{41}^0)}\,{\cal J}_2^-\ +
          {({\cal J}_{11}^0 + {\cal J}_{31}^0 - {\cal J}_{51}^0)}\,{\cal J}_3^-\ +
          {({\cal J}_{22}^0 + {\cal J}_{32}^0 - {\cal J}_{62}^0)}\,{\cal J}_3^- \bigg] \ - \
\\ &
   \frac{2}{m_2} \bigg[({\cal J}_{11}^0 + {\cal J}_{41}^0 - {\cal J}_{21}^0)\,{\cal J}_4^-\ +
          ({\cal J}_{11}^0 + {\cal J}_{51}^0 - {\cal J}_{31}^0)\,{\cal J}_5^-\ +
          ({\cal J}_{44}^0 + {\cal J}_{54}^0 - {\cal J}_{64}^0)\,{\cal J}_5^-
    \bigg] \ - \
\\ &
  \frac{2}{m_3} \bigg[({\cal J}_{22}^0 + {\cal J}_{42}^0 - {\cal J}_{12}^0)\,{\cal J}_4^-\ +
          ({\cal J}_{22}^0 + {\cal J}_{62}^0 - {\cal J}_{32}^0)\,{\cal J}_6^-\ +
          ({\cal J}_{44}^0 + {\cal J}_{64}^0 - {\cal J}_{54}^0)\,{\cal J}_6^-
    \bigg] \ - \
\\ &
  \frac{2}{m_4} \bigg[({\cal J}_{33}^0 + {\cal J}_{53}^0 - {\cal J}_{13}^0)\,{\cal J}_5^-\ +
          ({\cal J}_{33}^0 + {\cal J}_{63}^0 -{\cal J}_{23}^0)\,{\cal J}_6^-\ +
          ({\cal J}_{55}^0  +{\cal J}_{65}^0  - {\cal J}_{45}^0 )\,{\cal J}_6^-
    \bigg] \ + \
\\ &
   2\,\omega\,\bigg[ \,2\,a\,{\cal J}_{11}^0 \ + \   \,2\,b\,{\cal J}_{22}^0 \ + \ 2\,c\,{\cal J}_{33}^0 \ + \ \,2\,e\,{\cal J}_{44}^0  \ + \ \,2\,f\,{\cal J}_{55}^0 \ + \  \,2\,g\,{\cal J}_{66}^0    \ + \
\\ &
   \frac{\mu_{13}\,b\, \left({\cal J}_{11}^0+{\cal J}_{21}^0-{\cal J}_{41}^0\right)}{m_1}\ + \ \frac{\mu_{23}\,e \left({\cal J}_{11}^0-{\cal J}_{21}^0+{\cal J}_{41}^0\right)}{m_2} \ + \ \frac{\mu_{14}\,c \,\left({\cal J}_{11}^0+{\cal J}_{31}^0-{\cal J}_{51}^0\right)}{m_1} \ + \
\\ &
\frac{\mu_{24}\,f \left({\cal J}_{11}^0-{\cal J}_{31}^0+{\cal J}_{51}^0\right)}{m_2}  \ + \ \frac{\mu_{12}\,a\, \left({\cal J}_{12}^0+{\cal J}_{22}^0-{\cal J}_{42}^0\right)}{m_1}\ + \ \frac{\mu_{14}\,c \left({\cal J}_{22}^0-{\cal J}_{62}^0+{\cal J}_{32}^0\right)}{m_1} \ + \
\\ &
\frac{\mu_{23}\,e\, \left({\cal J}_{42}^0+{\cal J}_{22}^0-{\cal J}_{12}^0\right)}{m_3} \ + \ \frac{\mu_{34}\,g \left({\cal J}_{62}^0-{\cal J}_{32}^0+{\cal J}_{22}^0\right)}{m_3}   \ + \  \frac{\mu_{12}\,a \left({\cal J}_{33}^0+{\cal J}_{13}^0-{\cal J}_{53}^0\right)}{m_1}  \ + \
\\ &
 \frac{\mu_{13}\,b \left({\cal J}_{33}^0+{\cal J}_{23}^0-{\cal J}_{63}^0\right)}{m_1} \ +  \ \frac{\mu_{24}\,f \left({\cal J}_{33}^0+{\cal J}_{53}^0-{\cal J}_{13}^0\right)}{m_4} \ + \ \frac{\mu_{34}\,g \left({\cal J}_{33}^0+{\cal J}_{63}^0-{\cal J}_{23}^0\right)}{m_4}    \  + \
\\ &
  \frac{\mu_{12}\,a \left({\cal J}_{44}^0+{\cal J}_{14}^0-{\cal J}_{24}^0\right)}{m_2}    \ + \ \frac{\mu_{13}\,b \left({\cal J}_{44}^0+{\cal J}_{24}^0-{\cal J}_{14}^0\right)}{m_3}    \ + \  \frac{\mu_{24}\,f \left({\cal J}_{44}^0+{\cal J}_{54}^0-{\cal J}_{64}^0\right)}{m_2}  \ + \
\\ &
\frac{\mu_{34}\,g \left({\cal J}_{44}^0+{\cal J}_{64}^0-{\cal J}_{54}^0\right)}{m_3}  \ + \  \frac{\mu_{12}\,a \left({\cal J}_{55}^0 +{\cal J}_{15}^0 -{\cal J}_{35}^0 \right)}{m_2}  \ + \ \frac{\mu_{14}\,c \left({\cal J}_{55}^0 +{\cal J}_{35}^0 -{\cal J}_{15}^0 \right)}{m_4}  \ + \
\\ &
\frac{\mu_{23}\,e \left({\cal J}_{55}^0 +{\cal J}_{45}^0 -{\cal J}_{65}^0 \right)}{m_2}  \ + \
  \frac{\mu_{34}\,g \left({\cal J}_{55}^0 +{\cal J}_{65}^0 -{\cal J}_{45}^0 \right)}{m_4}   \ + \  \frac{\mu_{13}\,b \left({\cal J}_{66}^0+{\cal J}_{26}^0-{\cal J}_{36}^0\right)}{m_3}  \ + \
\\ &
 \frac{\mu_{14}\,c \left({\cal J}_{66}^0+{\cal J}_{36}^0-{\cal J}_{26}^0\right)}{m_4}  \ +  \ \frac{\mu_{23}\,e \left({\cal J}_{66}^0+{\cal J}_{46}^0-{\cal J}_{56}^0\right)}{m_3}  \ + \  \frac{\mu_{24}\,f \left({\cal J}_{66}^0+{\cal J}_{56}^0-{\cal J}_{46}^0\right)}{m_4}    \bigg] \ .
\end{aligned}
\end{equation}

It acts on (\ref{P6}) as a filtration. The above Lie-algebraic operator $h^{(es)}({\cal J})$ is the main object of study in the present consideration.

\section{Four-body harmonic oscillator: special cases}
\label{limitingcases}

For the 4-body harmonic oscillator with potential (\ref{V3-es}), the following physically important cases emerge:
\begin{itemize}
  \item Four equal masses
  \item atomic-like case when one mass is infinite but three other masses are equal
  \item molecular-like two-center case when two masses are infinite and two other masses are equal
  \item molecular-like three-center case when three masses are infinite
\end{itemize}

In the forthcoming subsections these cases will be analyzed in detail.

\subsection{Four particles of equal masses}

\subsubsection{ {\it Arbitrary spring constants}}

Let us take the eigenvalue problem (\ref{H-3-body}) with potential (\ref{V3-es}) and consider the case of four particles of equal masses, namely $m_1=m_2=m_3=m_4=m$, but different spring constants $\nu_{12},\nu_{13},\ldots, \nu_{34}>0$. The exactly solvable potential (\ref{V3-es}) becomes
\begin{equation}
\label{V3-es-meq}
\begin{aligned}
   V^{(4m)}\ &  = \ \frac{1}{2}\,m\,\om^2\,[\ (2 a^2+a (b+c+e+f)-b e-c f)\,\rho_{12} \ + \   (2 b^2+b (a+c+e+g)-a e-c g)\,\rho_{13}
\\ &
\ + \ (2 c^2+c (a+b+f+g)-b g-a f))\,\,\rho_{14} \ + \ (2 e^2+e (a+b+f+g)-a b- fg))\,\,\rho_{23}   \ + \
\\ &
(2 f^2+f (a+c+e+g)-a c- eg))\,\,\rho_{24}  \ + \ (2 g^2+g (b+c+e+f)-bc- ef))\,\,\rho_{34}  \ ] \ ,
\end{aligned}
\end{equation}
where we have used (\ref{freq-3}). This is a type of non-isotropic 6-body harmonic oscillator (in $r-$variables) with different spring constants.
In this case the ground state function (\ref{Psi03}) is reduced to
\begin{equation}
\label{Psi03-meq}
   \Psi_0^{(4m)}\ =\ e^{-\frac{\om\,m }{2}\,(\,a\,\rho_{12}\ +\ b\,\rho_{13}\ +\ c\,\rho_{14} \ +\ e\,\rho_{23}\ +\ f\,\rho_{24}\ +\ g\,\rho_{34}\,)}\ ,
\end{equation}
while its energy (\ref{E0en}) remains unchanged
\begin{equation*}
\label{e03-meq}
E_0^{(4m)}\ = \ \om \,d\,(a\,+\,b\,+\,c\,+\,e\,+\,f\,+\,g) \ .
\end{equation*}

The Lie-algebraic operator (\ref{H4LieA}) simplifies to
{\small
\begin{equation}
\label{}
\begin{aligned}
 h_{4m}^{(es)}({\cal J}) & \ = \ - \frac{4}{m}\,\bigg[\ {\cal J}_{11}^0\,{\cal J}_1^- \ + \  {\cal J}_{22}^0\,{\cal J}_2^-
     \ + \  {\cal J}_{33}^0\,{\cal J}_3^- \ + \    {\cal J}_{44}^0\,{\cal J}_4^- \ + \ {\cal J}_{55}^0\,{\cal J}_5^- \ + \
   {\cal J}_{66}^0\,{\cal J}_6^-  \,\bigg]
\\ &
    \ - \
\frac{2\,d}{m} \,\bigg[\,{\cal J}_1^- \ + \  {\cal J}_2^- \ + \ {\cal J}_3^- \ + \ {\cal J}_4^- \  + \ {\cal J}_5^- \  + \ {\cal J}_6^- \,\bigg] \ - \
\\ &
   \frac{2}{m} \bigg[{({\cal J}_{11}^0 + {\cal J}_{21}^0 - {\cal J}_{41}^0)}\,{\cal J}_2^-\ + \
          {({\cal J}_{11}^0 + {\cal J}_{31}^0 - {\cal J}_{51}^0)}\,{\cal J}_3^-\ + \
          {({\cal J}_{22}^0 + {\cal J}_{32}^0 - {\cal J}_{62}^0)}\,{\cal J}_3^- \ + \
\\ &
   ({\cal J}_{11}^0 + {\cal J}_{41}^0 - {\cal J}_{21}^0)\,{\cal J}_4^-\ +  \
          ({\cal J}_{11}^0 + {\cal J}_{51}^0 - {\cal J}_{31}^0)\,{\cal J}_5^-\ + \
          ({\cal J}_{44}^0 + {\cal J}_{54}^0 - {\cal J}_{64}^0)\,{\cal J}_5^-
   \ + \
\\ &
   ({\cal J}_{22}^0 + {\cal J}_{42}^0 - {\cal J}_{12}^0)\,{\cal J}_4^-\ + \
          ({\cal J}_{22}^0 + {\cal J}_{62}^0 - {\cal J}_{32}^0)\,{\cal J}_6^-\ + \
          ({\cal J}_{44}^0 + {\cal J}_{64}^0 - {\cal J}_{54}^0)\,{\cal J}_6^-
    \ + \
\\ &
 ({\cal J}_{33}^0 + {\cal J}_{53}^0 - {\cal J}_{13}^0)\,{\cal J}_5^-\ + \
          ({\cal J}_{33}^0 + {\cal J}_{63}^0 -{\cal J}_{23}^0)\,{\cal J}_6^-\ + \
          ({\cal J}_{55}^0  +{\cal J}_{65}^0  - {\cal J}_{45}^0 )\,{\cal J}_6^-
    \bigg] \ + \
\\ &
   \omega\,\bigg[ \,(4\,a+b+c+e+f)\,{\cal J}_{11}^0 \, + \,   (4\,b+a+c+e+g)\,{\cal J}_{22}^0 \ + \
(4\,c+a+b+f+g)\,{\cal J}_{33}^0 \, +\,
\\ &
 (4\,e+a+b+f+g)\,{\cal J}_{44}^0
 \, + \, (4\,f+a+c+e+g)\,{\cal J}_{55}^0 \, + \,  (4\,g+b+c+e+f)\,{\cal J}_{66}^0  \ + \
\\ &
 (b-e)({\cal J}_{21}^0-{\cal J}_{41}^0) \,+\, (c-f)({\cal J}_{31}^0-{\cal J}_{51}^0)
 \,+\, (a-e)({\cal J}_{12}^0-{\cal J}_{42}^0)   \,+\,
 (c-g)({\cal J}_{32}^0-{\cal J}_{62}^0)
\\ &
  \,+\, (a-f)({\cal J}_{13}^0-{\cal J}_{53}^0)
 \,+\, (b-g)({\cal J}_{23}^0-{\cal J}_{63}^0)
  \,+\, (a-b)({\cal J}_{14}^0-{\cal J}_{24}^0)  \,+\, (f-g)({\cal J}_{54}^0-{\cal J}_{64}^0)
\\ &
   \,+\, (a-c)({\cal J}_{15}^0-{\cal J}_{35}^0)
  \,+\,(e-g)({\cal J}_{45}^0-{\cal J}_{65}^0)
\,+\, (b-c)({\cal J}_{26}^0-{\cal J}_{36}^0)  \,+\, (e-f)({\cal J}_{46}^0-{\cal J}_{56}^0) \bigg] \ .
\end{aligned}
\end{equation}
}

\subsubsection{ {\it Equal spring constants}}

Now, let us take the case of four particles of equal masses and equal spring constants, namely $m_1=m_2=m_3=m_4=m$ and $a=b=c=e=f=g$. The potentials (\ref{V3-es}), (\ref{V3-es-meq}) degenerate to a type of six-dimensional isotropic 4-body harmonic oscillator without separation of  $\rho$-variables
\begin{equation}
\label{V3-es-meq-aeq}
   V^{(4a)}\ =\ 2\,m\,a^2\,\om^2\,(\rho_{12} \ + \ \rho_{13} \ + \ \rho_{14}\ + \ \rho_{23} \ + \ \rho_{24} \ + \ \rho_{34}) \ ,
\end{equation}
\cite{MTE:2018}.
In this case the wavefunction (\ref{Psi03}) and the corresponding ground state energy (\ref{E0en}) take the form
\begin{equation}
\label{Psi03-3a}
   \Psi_0^{(4a)}\ =\ e^{-\frac{\omega\,m}{2}\,a\,(\,\rho_{12} \ + \ \rho_{13} \ + \ \rho_{14}\ + \ \rho_{23} \ + \ \rho_{24} \ + \ \rho_{34}\,)}\ ,
\end{equation}
\begin{equation}
\label{E03-3a}
E_0^{(4a)}\ = \ 6\,\omega \,d\,a \ ,
\end{equation}
respectively. The Lie-algebraic operator (\ref{H4LieA}) simplifies to
\begin{equation}
\label{algebraic-meq-3a}
\begin{aligned}
 h_{4a}^{(es)}({\cal J}) & \ = \ - \frac{4}{m}\,\bigg[\ {\cal J}_{11}^0\,{\cal J}_1^- \ + \  {\cal J}_{22}^0\,{\cal J}_2^-
     \ + \  {\cal J}_{33}^0\,{\cal J}_3^- \ + \    {\cal J}_{44}^0\,{\cal J}_4^- \ + \ {\cal J}_{55}^0\,{\cal J}_5^- \ + \
   {\cal J}_{66}^0\,{\cal J}_6^-  \,\bigg]
\\ &
    \ - \
\frac{2\,d}{m} \,\bigg[\,{\cal J}_1^- \ + \  {\cal J}_2^- \ + \ {\cal J}_3^- \ + \ {\cal J}_4^- \  + \ {\cal J}_5^- \  + \ {\cal J}_6^- \,\bigg] \ - \
\\ &
   \frac{2}{m} \bigg[{({\cal J}_{11}^0 + {\cal J}_{21}^0 - {\cal J}_{41}^0)}\,{\cal J}_2^-\ + \
          {({\cal J}_{11}^0 + {\cal J}_{31}^0 - {\cal J}_{51}^0)}\,{\cal J}_3^-\ + \
          {({\cal J}_{22}^0 + {\cal J}_{32}^0 - {\cal J}_{62}^0)}\,{\cal J}_3^- \bigg] \ + \
\\ &
  ({\cal J}_{11}^0 + {\cal J}_{41}^0 - {\cal J}_{21}^0)\,{\cal J}_4^-\ +  \
          ({\cal J}_{11}^0 + {\cal J}_{51}^0 - {\cal J}_{31}^0)\,{\cal J}_5^-\ +  \
          ({\cal J}_{44}^0 + {\cal J}_{54}^0 - {\cal J}_{64}^0)\,{\cal J}_5^-
    \ + \
\\ &
 ({\cal J}_{22}^0 + {\cal J}_{42}^0 - {\cal J}_{12}^0)\,{\cal J}_4^-\ +  \
          ({\cal J}_{22}^0 + {\cal J}_{62}^0 - {\cal J}_{32}^0)\,{\cal J}_6^-\ +  \
          ({\cal J}_{44}^0 + {\cal J}_{64}^0 - {\cal J}_{54}^0)\,{\cal J}_6^-
     \ + \
\\ &
  ({\cal J}_{33}^0 + {\cal J}_{53}^0 - {\cal J}_{13}^0)\,{\cal J}_5^-\ + \
          ({\cal J}_{33}^0 + {\cal J}_{63}^0 -{\cal J}_{23}^0)\,{\cal J}_6^-\ +  \
          ({\cal J}_{55}^0  +{\cal J}_{65}^0  - {\cal J}_{45}^0 )\,{\cal J}_6^-
    \bigg] \ + \
\\ &
   8\,a\,\omega\,\bigg[ \,{\cal J}_{11}^0 \, + \,   {\cal J}_{22}^0 \ + \
{\cal J}_{33}^0 \, +\,{\cal J}_{44}^0 \, + \,{\cal J}_{55}^0 \, + \,  {\cal J}_{66}^0 \, \bigg] \ .
\end{aligned}
\end{equation}

The spectrum of (\ref{algebraic-meq-3a}) is the following:
\begin{equation}
\label{ep-3b-meq-3a}
  \varepsilon \ =\ 8\,a\,\omega\,(N_1+N_2+N_3+N_4+N_5+N_6)\ ,\
\end{equation}
where $N_1,N_2,\ldots,N_6=0,1,2,\ldots$ are quantum numbers.

It is worth mentioning that in the case of equal masses and equal spring constants there exists a nontrivial $sl(7)$-QES extension of (\ref{algebraic-meq-3a}) where all eigenfunctions are proportional to
\begin{equation}
\label{psiQES-3-equal-mass}
 \tilde {\Psi}_0 \ =\ \Psi_0^{(4a)} \times e^{-A(\,\rho_{12}^2\ + \ \rho_{13}^2\ + \ \rho_{14}^2 \ +\ \rho_{23}^2\ + \ \rho_{24}^2\ +\ \rho_{34}^2\,)} \ ,
\end{equation}
see (\ref{Psi03-3a}), $A>0$. Thus, the exponent in (\ref{psiQES-3-equal-mass}) is a second degree polynomial in $\rho$'s. The details are presented later in Section \ref{reduction}.

\subsection{Atomic case: $m_1=\infty$}

An interesting special case of the four-body oscillator emerges when $m_1 \rightarrow \infty$ and other three masses are kept equal $m_2=m_3=m_4=m$. In this case the potential (\ref{V3-es}) reduces to
\begin{equation}
\label{V3-es-at}
\begin{aligned}
 V^{(at)}\ &  = \ \frac{1}{2}m\,\omega^2\,\bigg[  \, 2(2 \,a^2 +a(e+f)-e\,b-c\,f  )\,\rho_{12} \ + \  2(2 \,b^2+b(e+g)-a\,e-c\,g   )\,\rho_{13} \ + \
\\ &
2(2 \,c^2+c(f+g)-a\,f-b\,g   )\,\rho_{14} \ + \  (2 \,e^2 +e(2a+2b+f+g)-f\,g  )\,\rho_{23} \ + \
\\ &
 (2 \,f^2 +f(2a+2c+e+g)-e\,g  )\,\rho_{24}  \ + \  (2 \,g^2 +g(2b+2c+e+f)-e\,f  )\,\rho_{34}   \bigg]\ ,
\end{aligned}
\end{equation}

c.f. (\ref{V3-es-meq}).
In general, the limit $m_1 \rightarrow \infty$ when keeping $m_{2,3,4}$ finite corresponds to physical atomic systems where one mass is much heavier than the others (for instance, as in the neutral Li atom or the positively charged lithium-like ions: $Be^+$, $B^{2+}$, etc). We call this case {\it atomic}.

For the atomic case the ground state function (\ref{Psi03}) is simplified to
\begin{equation}
\label{Psi03-at}
   \Psi_0^{(at)}\ =\ e^{-\frac{\omega\, m}{2}\,(2\,a\,\rho_{12}\ + \ 2\,b\,\rho_{13}\ +\ 2\,c\,\rho_{14}\ +\ e\,\rho_{23}\ +\ f\,\rho_{24}\ +\ g\,\rho_{34}\,)}\ .
\end{equation}
Since the general ground state energy (\ref{E0en}) does not depend on masses, in this case is given by
\begin{equation}
\label{e03-at}
E_0^{(at)}\  =  \ \omega \,d\,(a\,+\,b\,+\,c\,+\,e\,+\,f\,+\,g)   \ ,
\end{equation}
as well. Also, the Lie-algebraic operator (\ref{H-4body-algebraic}) becomes

\begin{equation}
\label{algebraic-atomic}
\begin{aligned}
 h_{at}^{(es)}({\cal J}) & \ = \ - 2\,\frac{1}{m}\bigg[\,{\cal J}_{11}^0\,{\cal J}_1^- \ + \  {\cal J}_{22}^0\,{\cal J}_2^-
     \ + \  {\cal J}_{33}^0\,{\cal J}_3^- \ + \    2\,{\cal J}_{44}^0\,{\cal J}_4^- \ + \ 2\,{\cal J}_{55}^0\,{\cal J}_5^- \ + \  2\,{\cal J}_{66}^0\,{\cal J}_6^-  \,\bigg]
\\ &
    \quad - \
\frac{d}{m} \,\bigg[\,{\cal J}_1^- + {\cal J}_2^- + {\cal J}_3^- + 2\,{\cal J}_4^-  + 2\,{\cal J}_5^-  + 2\,{\cal J}_6^- \,\bigg] \ - \
\\ &
   \frac{2}{m} \bigg[({\cal J}_{11}^0 + {\cal J}_{41}^0 - {\cal J}_{21}^0)\,{\cal J}_4^-\ + \
          ({\cal J}_{11}^0 + {\cal J}_{51}^0 - {\cal J}_{31}^0)\,{\cal J}_5^-\ + \
          ({\cal J}_{44}^0 + {\cal J}_{54}^0 - {\cal J}_{64}^0)\,{\cal J}_5^-
    \ + \
\\ &
  ({\cal J}_{22}^0 + {\cal J}_{42}^0 - {\cal J}_{12}^0)\,{\cal J}_4^-\ + \
          ({\cal J}_{22}^0 + {\cal J}_{62}^0 - {\cal J}_{32}^0)\,{\cal J}_6^-\ + \
          ({\cal J}_{44}^0 + {\cal J}_{64}^0 - {\cal J}_{54}^0)\,{\cal J}_6^-
     \ + \
\\ &
   ({\cal J}_{33}^0 + {\cal J}_{53}^0 - {\cal J}_{13}^0)\,{\cal J}_5^-\ + \
          ({\cal J}_{33}^0 + {\cal J}_{63}^0 -{\cal J}_{23}^0)\,{\cal J}_6^-\ + \
          ({\cal J}_{55}^0  +{\cal J}_{65}^0  - {\cal J}_{45}^0 )\,{\cal J}_6^-
    \bigg] \ + \
\\ &
   \omega\,\bigg[ \,(4\,a\,+\,e\,+\,f)\,{\cal J}_{11}^0 \ + \   \,(4\,b\,+\,e\,+\,g)\,{\cal J}_{22}^0 \ + \ (4\,c\,+\,f\,+\,g)\,{\cal J}_{33}^0 \ + \
\\ &
   \,(4\,e+2a+2b+f+g)\,{\cal J}_{44}^0  \ + \ \,(4\,f+2a+2c+e+g)\,{\cal J}_{55}^0 \ + \  \,(4\,g+2b+2c+e+f)\,{\cal J}_{66}^0    \ + \
\\ &
2\,a \left({\cal J}_{15}^0  +{\cal J}_{14}^0-{\cal J}_{35}^0-{\cal J}_{24}^0\right)    \ + \ 2\,b \left({\cal J}_{26}^0+{\cal J}_{24}^0-{\cal J}_{36}^0-{\cal J}_{14}^0\right) \ + \
\\ &
 2\,c \left({\cal J}_{36}^0+{\cal J}_{35}^0-{\cal J}_{26}^0 -{\cal J}_{15}^0 \right)  \ + \ e\, \left({\cal J}_{45}^0+{\cal J}_{41}^0 +{\cal J}_{42}^0+{\cal J}_{46}^0-{\cal J}_{12}^0-{\cal J}_{56}^0-{\cal J}_{21}^0-{\cal J}_{65}^0\right)  \ + \
\\ &
f \left({\cal J}_{56}^0+{\cal J}_{51}^0+{\cal J}_{54}^0+{\cal J}_{53}^0-{\cal J}_{64}^0-{\cal J}_{13}^0-{\cal J}_{31}^0-{\cal J}_{46}^0\right)
\  + \
\\ &
g \left({\cal J}_{65}^0  +{\cal J}_{64}^0+{\cal J}_{62}^0+{\cal J}_{63}^0-{\cal J}_{54}^0-{\cal J}_{32}^0-{\cal J}_{45}^0-{\cal J}_{23}^0\right)
       \, \bigg] \ .
\end{aligned}
\end{equation}

\subsubsection{Laplace-Beltrami operator: underlying geometry}

For the atomic case, $m_1\rightarrow \infty$, the co-metric defined by the coefficients in front of second derivatives in (\ref{algebraic-atomic})
{\small
\begin{equation}
\label{gmn33-rhoatomic}
 g^{\mu \nu}_{(at)}(\rho)\ = \left(
\begin{array}{cccccc}
 \frac{2}{m}\, \rho _{12} & 0 & 0 & \frac{\rho _{12}-\rho _{13}+\rho _{23}}{m} & \frac{\rho _{12}-\rho _{14}+\rho _{24}}{m} & 0 \\
 0& \frac{2}{m}\, \rho _{13} & 0& \frac{\rho _{13}+\rho _{23}-\rho _{12}}{m} & 0 & \frac{\rho _{13}-\rho _{14}+\rho _{34}}{m} \\
 0 & 0 & \frac{2}{m} \,\rho _{14} & 0 & \frac{\rho _{14}+\rho _{24}-\rho _{12}}{m} & \frac{\rho _{14}+\rho _{34}-\rho _{13}}{m} \\
 \frac{\rho _{12}-\rho _{13}+\rho _{23}}{m} & \frac{\rho _{13}+\rho _{23}-\rho _{12}}{m} & 0 & \frac{4}{m} \,\rho _{23} & \frac{\rho _{23}+\rho _{24}-\rho _{34}}{m} & \frac{\rho _{23}-\rho _{24}+\rho _{34}}{m} \\
 \frac{\rho _{12}-\rho _{14}+\rho _{24}}{m} & 0 & \frac{\rho _{14}+\rho _{24}-\rho _{12}}{m} & \frac{\rho _{23}+\rho _{24}-\rho _{34}}{m} & \frac{4}{m}\, \rho _{24} & \frac{\rho _{24}+\rho _{34}-\rho _{23}}{m} \\
 0 & \frac{\rho _{13}-\rho _{14}+\rho _{34}}{m} & \frac{\rho _{14}+\rho _{34}-\rho _{13}}{m} & \frac{\rho _{23}-\rho _{24}+\rho _{34}}{m} & \frac{\rho _{24}+\rho _{34}-\rho _{23}}{m} & \frac{4}{m}\, \rho _{34} \\
\end{array}
\right) \ ,
\end{equation}
}
is proportional to $1/m$ and possesses a factorizable determinant
\[
 D_{(at)}\ \equiv \ {\rm Det} g^{\mu \nu}_{(at)}\ = \ \frac{9216}{m^6}\,V_4^2 \ \times
\]
\begin{equation}
\label{gmn33-rho-det-Matomic}
 \bigg[\, ( \rho_{12}\, + \, \rho_{13} \, + \, \rho_{14}\,)(\, S^2(\sqrt{\rho_{13}},\,\sqrt{\rho_{14}},\,\sqrt{\rho_{34}})\ + \  S^2(\sqrt{\rho_{12}},\,\sqrt{\rho_{14}},\,\sqrt{\rho_{24}})\ + \ S^2(\sqrt{\rho_{12}},\,\sqrt{\rho_{13}},\,\sqrt{\rho_{23}})) \ - \ 9\,V_4^2  \,\bigg]                     \ ,
\end{equation}
which is positive definite (cf. (\ref{gmn33-rho-det-M})). We emphasize that the operator (\ref{algebraic-atomic}) is six-dimensional, all six $\rho$-variables remain dynamical.

Next, making the gauge transformation
\begin{equation}
         \Gamma_{(at)}^{-1}\,\Psi_0^{(at)}\, [\,h_{at}^{(es)}\ + \ E_0^{(at)}\,]\,{(\Psi_0^{(at)})}^{-1}\,\Gamma_{(at)} \ = \ \Gamma_{(at)}^{-1}\, (\,-\Delta_{\rm rad}^{(at)}\,)\,\Gamma_{(at)} \ = \  -\Delta^{(at)}_{LB}(\rho) \ + \ V^{(at)}_{\rm eff} \ ,
\end{equation}
with determinant (\ref{gmn33-rho-det-Matomic}) as the factor,
\begin{equation}
         \Gamma_{(at)} \ = \  D_{(at)}^{-\frac{1}{4}}\,V_4^{1 - \frac{d}{4}} \quad ,
\end{equation}
we obtain a Laplace-Beltrami operator $\Delta^{(at)}_{LB}(\rho)$, with metric (\ref{gmn33-rhoatomic}), plus the effective potential
\begin{equation}
V^{(at)}_{\rm eff} \ \equiv \ \lim_{m_1\rightarrow \infty} V_{\rm eff} \ ,
\end{equation}
where $V_{\rm eff}$ is given by (\ref{Veffgen}}). Hence, from the original Hamiltonian (\ref{Hgen}) we arrive at the spectral problem for the Schr\"odinger operator
\begin{equation}\label{}
  H^{(at)}_{LB} \ = \  -\Delta^{(at)}_{LB} \ + \ V^{(at)}_{\rm eff} \ + \ V \ ,
\end{equation}
in the $\rho$-space with $d > 2$. The $d$-independent Laplace-Beltrami operator $-\Delta_{LB}^{(at)}$ plays the role of a kinetic energy again.

\subsection{Molecular two-center case: $m_1,m_2=\infty$}

In the molecular-like case two masses are considered infinitely heavy, $m_{1,2} \rar \infty$, thus, the reduced mass $\mu_{12}$ also tends to infinity, while the two masses $m_3,m_4=m$ remain finite. Sometimes it is called the Born-Oppenheimer approximation of zero order. It implies that the coordinate $\rho_{12}$ is classical (thus, unchanged in dynamics, being constant of motion in the process of evolution, it can be treated as external parameter), while five other $\rho$-variables remain dynamical. The 4-body problem is converted into a two-body system in a two-center potential. Given the ground state function (\ref{Psi03}), the harmonic potential (\ref{V3-es}) depends on masses via (\ref{freq-3}). In order to keep the potential finite in the molecular limit we set the parameter $a=0$, the coefficient on front of $\rho_{12}$ in (\ref{Psi03}), from the very beginning.

In this case the potential (\ref{V3-es}) reduces to \footnote{We must emphasize that the potential (\ref{V3-es-mol}) is defined up to an additive constant, which can depend on classical coordinate $\rho_{12}$. This constant defines the reference point for energy.}
\begin{equation}
\label{V3-es-mol}
\begin{aligned}
 V^{(mol)}\ &  = \ m\,\omega^2\,\bigg[  \,(2 b^2+b (2 e+g)-c g)\,\rho_{13} \ + \  (2 c^2+c (2 f+g)-b g )\,\rho_{14} \ + \
\\ &
( 2 e^2+e(2b + g)-f g   )\,\rho_{23} \ + \  (2 f^2+f(2c+ g)-e g)\,\rho_{24}  + \  g (b+c+e+f+g) \,\rho_{34}   \bigg]\ ,
\end{aligned}
\end{equation}

c.f. (\ref{V3-es-meq}), (\ref{V3-es-at}). In general, the limit $m_1,m_2 \rightarrow \infty$ when keeping $m_{3,4}$ finite corresponds to physical molecular systems where two masses are much heavier than the others (for instance, as the $H_2$ molecule). We call this {\it molecular two-center case}.

For the molecular two-center case the ground state function (\ref{Psi03}) is simplified to
\begin{equation}
\label{Psi03-mol}
   \Psi_0^{(mol)}\ =\ e^{-\frac{\omega\, m}{2}\,(\, 2\,b\,\rho_{13}\ +\ 2\,c\,\rho_{14}\ +\ 2\,e\,\rho_{23}\ +\ 2\,f\,\rho_{24}\ +\ g\,\rho_{34}\,)}\ ,
\end{equation}
and the ground state energy reads
\begin{equation}
\label{e03-mol}
E_0^{(mol)}\  =  \ \omega \,d\,(b\,+\,c\,+\,e\,+\,f\,+\,g)  \ + \ 2\,m\,\omega^2\,(b\,e+c\,f)\,\rho_{12}  \ .
\end{equation}

Note that $E_0^{(mol)}$ is measured from the reference point $V^{(mol)}(0)=0$ and it is always larger than (or equal to) the exact energy (\ref{E0en}) at $a=0$, $E_0^{(mol)} \geq E_0^{(a=0)}$. $E_0^{(mol)}$ takes minimal value at $\rho_{12}=0$, where it coincides with exact ground state energy $E_0^{(a=0)}$.\\

In the limit $m_{1,2} \rar \infty$ the radial Laplacian (\ref{addition3-3r-M}) (as well as the associated Laplace-Beltrami operator) loses the property of invertability: both 1st row and 1st column in (\ref{gmn33-rho}) vanish as well as the determinant of cometric $D_m$ (\ref{gmn33-rho-det-M}). However, the Lie-algebraic operator (\ref{H4LieA}) at $a=0$ in the molecular limit is well-defined and finite. It reads,

\begin{equation}
\label{algebraic-molecular}
\begin{aligned}
 h_{mol}^{(es)}({\cal J}) & \ = \ - 2\,\frac{1}{m}\bigg[\,  {\cal J}_{22}^0\,{\cal J}_2^-
     \ + \  {\cal J}_{33}^0\,{\cal J}_3^- \ + \    {\cal J}_{44}^0\,{\cal J}_4^- \ + \ {\cal J}_{55}^0\,{\cal J}_5^- \ + \  2\,{\cal J}_{66}^0\,{\cal J}_6^-  \,\bigg]
\\ &
    \quad - \
\frac{d}{m} \,\bigg[\,{\cal J}_1^- + {\cal J}_2^- + {\cal J}_3^- + 2\,{\cal J}_4^-  + 2\,{\cal J}_5^-  + 2\,{\cal J}_6^- \,\bigg] \ - \
\\ &
 \frac{2}{m} \,\bigg[\, ({\cal J}_{22}^0 + {\cal J}_{42}^0 - {\cal J}_{12}^0)\,{\cal J}_4^-\ + \
          ({\cal J}_{22}^0 + {\cal J}_{62}^0 - {\cal J}_{32}^0)\,{\cal J}_6^-\ + \
          ({\cal J}_{44}^0 + {\cal J}_{64}^0 - {\cal J}_{54}^0)\,{\cal J}_6^-
     \ + \
\\ &
   ({\cal J}_{33}^0 + {\cal J}_{53}^0 - {\cal J}_{13}^0)\,{\cal J}_5^-\ + \
          ({\cal J}_{33}^0 + {\cal J}_{63}^0 -{\cal J}_{23}^0)\,{\cal J}_6^-\ + \
          ({\cal J}_{55}^0  +{\cal J}_{65}^0  - {\cal J}_{45}^0 )\,{\cal J}_6^-
    \bigg] \ + \
\\ &
   \omega\,\bigg[ \,    \,(4\,b\,+\,2e\,+\,g)\,{\cal J}_{22}^0 \ + \ (4\,c\,+\,2f\,+\,g)\,{\cal J}_{33}^0 \ + \ (4\,e\,+\,2b\,+\,g)\,{\cal J}_{44}^0  \ + \ \,(4\,f\,+\,2c\,+\,g)\,{\cal J}_{55}^0 \ + \
\\ &
     \,(4\,g+2b+2c+2e+2f)\,{\cal J}_{66}^0    \ + \
 2\,b \left({\cal J}_{26}^0+{\cal J}_{24}^0-{\cal J}_{36}^0-{\cal J}_{14}^0\right) \ + \ 2\,c \left({\cal J}_{36}^0+{\cal J}_{35}^0-{\cal J}_{26}^0 -{\cal J}_{15}^0 \right)
\\ &
\ + \ 2\,e\, \left({\cal J}_{42}^0+{\cal J}_{46}^0-{\cal J}_{12}^0-{\cal J}_{56}^0\right)  \ + \ 2\,f \left({\cal J}_{56}^0{\cal J}_{54}^0-{\cal J}_{64}^0-{\cal J}_{46}^0\right)
\\ &
\  + \ g\, \left({\cal J}_{65}^0  +{\cal J}_{64}^0+{\cal J}_{62}^0+{\cal J}_{63}^0-{\cal J}_{54}^0-{\cal J}_{32}^0-{\cal J}_{45}^0-{\cal J}_{23}^0\right)
       \, \bigg] \ .
\end{aligned}
\end{equation}

Its eigenfunctions $\phi_{k_1,k_2,k_3,k_4,k_5}$ are marked by five integer (quantum) numbers $k_1,k_2,\ldots,k_5=0,1,\ldots$. They are multivariate polynomials in dynamical variables $\rho_{13},\rho_{14},\rho_{23},\rho_{24},\rho_{34}$ and its spectrum $\varepsilon_{k_1,k_2,k_3,k_4,k_5}$ is linear in quantum numbers $(k_1,k_2,k_3,k_4,k_5)$ and $\varepsilon_{0,0,0,0,0}=0$. At $a=0$, in the molecular limit the eigenvalue problem (\ref{H-3-body}) becomes

\begin{equation}
\label{Hmole}
  {\cal H}^{(mol)}\,\Psi(\rho)  \ \equiv \    \big[\,-\Delta^{(mol)}_{\rm rad}(\rho)\ + \ V^{(mol)}(\rho)\,\big]\,\Psi(\rho)\ =\ E \,\Psi(\rho)\ ,\
\end{equation}

where

\begin{equation}
        -\Delta^{(mol)}_{\rm rad}   \ \equiv \  \Psi_0^{(mol)}\, [\,h_{mol}^{(es)}\ + \ E_0^{(mol)}\,]\,{(\Psi_0^{(mol)})}^{-1} \ .
\end{equation}

From the viewpoint of the Born-Oppenheimer approximation, widely used in molecular physics, where the particles $m_3,m_4$ can be associated with two electrons, respectively, and the whole system with a two-electron diatomic molecule, the Hamiltonian ${\cal H}^{(mol)}$ (\ref{Hmole}) has the meaning of the so-called electronic Hamiltonian. It describes the electronic degrees of freedom of a molecular system. The ground state energy $E_0^{(mol)}$ (\ref{e03-mol}) is called the {\it ground state energy potential curve} or, simply speaking, the potential curve.

The spectrum of the Hamiltonian (\ref{Hmole}) is
\begin{equation}
\label{Ek-mol}
  E^{(mol)}_{k_1,k_2,k_3,k_4,k_5} \ = \ E_0^{(mol)} + \varepsilon_{k_1,k_2,k_3,k_4,k_5} \ .
\end{equation}

\subsubsection{Laplace-Beltrami operator: underlying geometry}

For the molecular case, $m_1,m_2\rightarrow \infty$, the co-metric defined by the coefficients in front of second derivatives in (\ref{algebraic-molecular})
{\small
\begin{equation}
\label{gmn33-rhomolecular}
 g^{\mu \nu}_{(mol)}(\rho)\ = \left(
\begin{array}{cccccc}
  \frac{2}{m}\, \rho _{13} & 0& \frac{\rho _{13}+\rho _{23}-\rho _{12}}{m} & 0 & \frac{\rho _{13}-\rho _{14}+\rho _{34}}{m} \\
  0 & \frac{2}{m} \,\rho _{14} & 0 & \frac{\rho _{14}+\rho _{24}-\rho _{12}}{m} & \frac{\rho _{14}+\rho _{34}-\rho _{13}}{m} \\
  \frac{\rho _{13}+\rho _{23}-\rho _{12}}{m} & 0 & \frac{2}{m} \,\rho _{23} & 0 & \frac{\rho _{23}-\rho _{24}+\rho _{34}}{m} \\
  0 & \frac{\rho _{14}+\rho _{24}-\rho _{12}}{m} & 0 & \frac{2}{m}\, \rho _{24} & \frac{\rho _{24}+\rho _{34}-\rho _{23}}{m} \\
  \frac{\rho _{13}-\rho _{14}+\rho _{34}}{m} & \frac{\rho _{14}+\rho _{34}-\rho _{13}}{m} & \frac{\rho _{23}-\rho _{24}+\rho _{34}}{m} & \frac{\rho _{24}+\rho _{34}-\rho _{23}}{m} & \frac{4}{m}\, \rho _{34} \\
\end{array}
\right) \ ,
\end{equation}
}
is proportional to $1/m$ and possesses a factorizable determinant
\[
 D_{(mol)}\ \equiv \ {\rm Det} g^{\mu \nu}_{(mol)}\ = \ \frac{288}{m^5}\,V_4^2 \ \times
\]
\begin{equation}
\label{gmn33-rho-det-Mmolecular}
 \bigg[\, 2\,\rho_{12}\,(\, \rho_{13} \,+\,\rho_{14} \,+\,\rho_{23} \,+\,\rho_{24}\,-\,\rho_{12}\,) \ - \ {(\rho_{13}\,-\,\rho_{23})}^2 \ - \ {(\rho_{14}\,-\,\rho_{24})}^2  \,\bigg]   \ \equiv \   \frac{288}{m^5}\,V_4^2\,G_2                \ .
\end{equation}
We emphasize that the operator (\ref{algebraic-molecular}) is not six but five-dimensional. Here $\rho_{12}$ is a classical variable, it enters as a parameter only (it is not dynamical). As in the previous cases, making the gauge transformation
\begin{equation}
         \Gamma_{(mol)}^{-1}\,\Psi_0^{(mol)}\, [\,h_{mol}^{(es)}\ + \ E_0^{(mol)}\,]\,{(\Psi_0^{(mol)})}^{-1}\,\Gamma_{(mol)} \ = \ \Gamma_{(mol)}^{-1}\, (\,-\Delta_{\rm rad}^{(mol)}\,)\,\Gamma_{(mol)} \ = \  -\Delta^{(mol)}_{LB} \ + \ V^{(mol)}_{\rm eff} \ ,
\end{equation}
with the factor,
\begin{equation}
         \Gamma_{(mol)} \ = \  G_2^{-\frac{1}{4}}\,V_4^{\frac{3-d}{2}} \quad ,
\end{equation}
we obtain the Laplace-Beltrami operator $\Delta^{(mol)}_{LB}(\rho)$, with metric (\ref{gmn33-rhomolecular}), plus the effective potential
\begin{equation}
V^{(mol)}_{\rm eff} \ \equiv \  \frac{3\,\rho_{12}}{2\,m\,G_2} \ + \ \frac{(d-5)(d-3)\,G_2}{1152\,m\,V_4^2}     \ .
\end{equation}
Hence, from the original Hamiltonian (\ref{Hgen}) we arrive at the spectral problem for the Schr\"odinger operator
\begin{equation}\label{}
  H^{(mol)}_{LB} \ = \  -\Delta^{(mol)}_{LB} \ + \ V^{(mol)}_{\rm eff} \ + \ V \ ,
\end{equation}
in the $\rho$-space with $d > 2$.

\subsubsection{Molecular two-center case in the Born-Oppenheimer approximation: $m_1,m_2 \gg m_3,m_4$}

\bigskip

The 4-body oscillator model we deal with allows exact solvability and a critical analysis of the Born-Oppenheimer approximation as described below.

In the formalism of the Born-Oppenheimer approximation, the energy $E^{(mol)}_{k_1,k_2,k_3,k_4,k_5}$ (\ref{Ek-mol}) of the electronic Hamiltonian $ {\cal H}^{(mol)}$ (\ref{Hmole}) should appear as the potential in a two-body {\it nuclear} Hamiltonian.
In order to derive this Hamiltonian we should replace in (\ref{Hgen}) the sum of the kinetic energy of the bodies 3 and 4 by (\ref{Ek-mol}), separate center-of-mass of the first + second bodies and introduce the Euler coordinates
\[
   ({\bf r}_1\ ,\ {\bf r}_2) \rar ({\bf R}_{0}\ ,\ \rho=r_{12}^2\ ,\ \Om_{12})\ ,
\]
and then separate out $(d-1)$ angular variables $\{ \Om_{12} \}$. As a result
we arrive at
\begin{equation}
\label{Hnucl}
\begin{aligned}
 {\cal H}^{(nucl)}\ & =\ -\frac{1}{\mu}
    \bigg(2 \,\rho\, \pa^2_{\rho}\ +\ d\,\pa_{\rho} \bigg)\ +\ \frac{L (L+d-2)}{\rho}\ + \  2\,\omega^2\,(m\,b\,e+m\,c\,f+\nu_{12})\,\rho_{12} \ + \
\\ &
     \omega \,d\,(b\,+\,c\,+\,e\,+\,f\,+\,g)   \ +\ \varepsilon_{k_1,k_2,k_3,k_4,k_5}\ ,
\end{aligned}
\end{equation}
where $\mu \equiv  \frac{m_1\, m_2}{m_1+m_2}$ is the reduced mass, $m_{1,2}$ are masses of the nuclei, $\nu_{12}$ given by (\ref{freq-3}) and $L$ is its two-body nuclear angular momentum and $m=m_3=m_4$. Now $\rho \equiv \rho_{12}$ is restored as a dynamical variable.
For simplicity let us consider the ground state, putting $L=0$ and $k_1=k_2=k_3=k_4=k_5=0$ in the Hamiltonian (\ref{Hnucl}),
\begin{equation}
\label{Hnucl-0}
 {\cal H}_0^{(nucl)}\  =\ -\frac{1}{\mu}
    \bigg(2 \,\rho\, \pa^2_{\rho}\ +\ d\,\pa_{\rho} \bigg)\ +\  2\,\omega^2\,(m\,b\,e+m\,c\,f+\nu_{12})\,\rho \ + \ \omega \,d\,(b\,+\,c\,+\,e\,+\,f\,+\,g)   \ .
\end{equation}

This nuclear Hamiltonian defines the so-called vibrational spectrum of the ground state, its lowest eigenvalue (sometimes called zero-point energy) is
\begin{equation}
\label{Enucl-0}
    E_0^{(nucl)}\ =\ \left(\om\,d\,(b\,+\,c\,+\,e\,+\,f\,+\,g) \, + \,\om\,d\, \sqrt{\frac{(b\,e+c\,f)\,m}{\mu}\bigg(1 \,+\, \frac{\nu_{12}}{(b\,e+c\,f)\,m}\bigg)} \,\right)\ .
\end{equation}
Making comparison of the exact energy $E_0$ (\ref{E0en}) with $E_0^{(nucl)}$ we get
\begin{equation}
\label{del-Enucl-1}
   E_0^{(nucl)}\ -\ E_0 \ = \ \om\,d\,\bigg\{\sqrt{\frac{(b\,e+c\,f)\,m}{\mu}\bigg(1 \,+\, \frac{\nu_{12}}{(b\,e+c\,f)\,m}\bigg)} \ - \ a\,\bigg\} \ .
\end{equation}
This difference ``measures" the accuracy of the Born-Oppenheimer approximation: it tends to zero as $\mu \rar \infty$. For the relevant physical case $m_1=m_2=1$, we obtain the following expansion in powers of the small parameter $m\ll 1$
\begin{equation}
\label{del-Enucl-2}
   E_0^{(nucl)}\ -\ E_0 \ = \ \frac{\omega\,d}{2}\,\bigg(    (b+c+e+f)\,m \ - \ \frac{4 [a\, (b+c+e+f)-4 \,(b e+c f)]+{(b+c+e+f)}^2}{4\, a}\,m^2 \ + \ \ldots    \bigg) \ .
\end{equation}

As previously observed in the one-dimensional case $d=1$ for the three-body case \cite{Fernandez}, the Born-Oppenheimer approximation yields the leading term of the expansion of the exact result in powers of $m$ (or the ratio of the electron to nuclear mass) for any $d>1$.

\subsection{Molecular three-center case : $m_1,m_2,m_3=\infty$}

In the three-center case three masses are considered infinitely heavy, $m_{1,2,3} \rar \infty$, hence, the reduced masses $\mu_{12}$, $\mu_{13}$ and $\mu_{23}$ tends to infinity as well. The mass $m_4$ remains finite only. Sometimes it is called the Born-Oppenheimer approximation of zero order. It implies that the coordinates $\rho_{12}$, $\rho_{13}$, $\rho_{23}$ are classical and they can be treated as external parameters. The three other $\rho$-variables remain dynamical. The 4-body problem is converted into a one-body system in a three-center potential. In (\ref{Psi03}), from the very beginning we set the parameters $a=b=e=0$ so that the harmonic potential remains finite in the limit $m_{1,2,3} \rar \infty$.

In this case the potential (\ref{V3-es}) reduces to\footnote{The potential (\ref{V3-es-3cc}) is defined up to an additive constant, which can depend on classical coordinates $\rho_{12}$, $\rho_{13}$, $\rho_{23}$. This constant defines the reference point for energy.}
\begin{equation}
\label{V3-es-3cc}
\begin{aligned}
 V^{\rm (3-center)}\ &  = \ 2\,m\,(c+f+g)\,\omega^2\,\bigg[  \,  c\, \,\rho_{14} \ + \  f\,\rho_{24}  + \ g\, \rho_{34}   \bigg]\ ,
\end{aligned}
\end{equation}

c.f. (\ref{V3-es-meq}), (\ref{V3-es-at}), (\ref{V3-es-mol}). In general, the limit $m_1,m_2,m_3 \rightarrow \infty$ when keeping $m_{4}$ finite corresponds to physical molecular 4-body systems where three masses are much heavier than the fourth mass. We call this {\it molecular three-center one electron case}.

For the 3-center case the ground state function (\ref{Psi03}) is simplified to
\begin{equation}
\label{Psi03-3cc}
   \Psi_0^{\rm(3-center)}\ =\ e^{-\omega\, m\,(\,c\,\rho_{14}\ +\  f\,\rho_{24}\ +\ g\,\rho_{34}\,)}\ ,
\end{equation}
and the general ground state energy does not depend on masses, it reads
\begin{equation}
\label{e03-3cc}
E_0^{\rm(3-center)}\  =  \ \omega \,d\,(c\,+\,f\,+\,g)  \ + \ 2\,m\,\omega^2\,(c\,f\,\rho_{12}\,+\,c\,g\,\rho_{13}\,+\,f\,g\,\rho_{23}   )  \ .
\end{equation}

Note that $E_0^{\rm(3-center)}$ is measured from the reference point $V^{\rm(3-center)}(0)=0$ and it is always larger than (or equal to) the exact energy (\ref{E0en}) at $a=b=e=0$, $E_0^{\rm(3-center)} \geq E_0^{(a=b=e=0)}$. $E_0^{\rm(3-center)}$ takes minimal value at $\rho_{12}=\rho_{13}=\rho_{23}=0$, where it coincides with exact ground state energy $E_0^{(a=b=e=0)}$.\\

In the limit $m_{1,2,3} \rar \infty$ the radial Laplacian (\ref{addition3-3r-M}) (as well as the associated Laplace-Beltrami operator) loses the property of invertability: 1st, 2nd and 4th rows and columns in (\ref{gmn33-rho}) vanish as well as the determinant of cometric $D_m$ (\ref{gmn33-rho-det-M}). However, the Lie-algebraic operator (\ref{H4LieA}) at $a=b=e=0$ in the 3-center case limit is well-defined and finite. It reads,

\begin{equation}
\label{algebraic-3cc}
\begin{aligned}
 h_{\rm 3-center}^{(es)}({\cal J}) & \ = \ - 2\,\frac{1}{m}\bigg[\,   {\cal J}_{33}^0\,{\cal J}_3^- \ + \    {\cal J}_{55}^0\,{\cal J}_5^- \ + \  {\cal J}_{66}^0\,{\cal J}_6^-  \,\bigg]\ - \ \frac{d}{m} \,\bigg[\, {\cal J}_3^- \ + \ {\cal J}_5^- \ + \ {\cal J}_6^- \,\bigg] \ - \
\\ &
 \frac{2}{m} \,\bigg[\,
   ({\cal J}_{33}^0 + {\cal J}_{53}^0 - {\cal J}_{13}^0)\,{\cal J}_5^-\ + \
          ({\cal J}_{33}^0 + {\cal J}_{63}^0 -{\cal J}_{23}^0)\,{\cal J}_6^-\ + \
          ({\cal J}_{55}^0  +{\cal J}_{65}^0  - {\cal J}_{45}^0 )\,{\cal J}_6^-
    \bigg] \ + \
\\ &
   2\,\omega\,\bigg[ \,  (2\,c\,+\,f\,+\,g)\,{\cal J}_{33}^0   \ + \ (2\,f\,+\,c\,+\,g)\,{\cal J}_{55}^0 \ + \
    (2\,g\,+\,f\,+\,c)\,{\cal J}_{66}^0
\\ &
      \ + \  c\,({\cal J}_{35}^0\,+\,{\cal J}_{36}^0\,-\,{\cal J}_{15}^0\,-\,{\cal J}_{26}^0)
      \ + \  f\,({\cal J}_{53}^0\,+\,{\cal J}_{56}^0\,-\,{\cal J}_{13}^0\,-\,{\cal J}_{46}^0)
      \ + \
\\ &
       g\,({\cal J}_{63}^0\,+\,{\cal J}_{65}^0\,-\,{\cal J}_{23}^0\,-\,{\cal J}_{45}^0)
       \bigg] \ .
\end{aligned}
\end{equation}

Its eigenfunctions $\phi_{k_1,k_2,k_3}$ are marked by three integer (quantum) numbers $k_1,k_2,k_3$. They are multivariate polynomials in dynamical variables $\rho_{14},\rho_{24},\rho_{34}$ and its spectrum $\varepsilon_{k_1,k_2,k_3}$ is linear in quantum numbers $(k_1,k_2,k_3)$ and $\varepsilon_{0,0,0}=0$. At $a=b=e=0$, in the three-center limit the eigenvalue problem (\ref{H-3-body}) becomes

\begin{equation}
\label{H3cc}
  {\cal H}^{\rm(3-center)}\,\Psi(\rho)  \ \equiv \    \big[\,-\Delta^{\rm(3-center)}_{\rm rad}(\rho)\ + \ V^{\rm(3-center)}(\rho)\,\big]\,\Psi(\rho)\ =\ E \,\Psi(\rho)\ ,\
\end{equation}

where

\begin{equation}
        -\Delta^{\rm(3-center)}_{\rm rad}   \ \equiv \  \Psi_0^{\rm(3-center)}\, [\,h_{\rm3-center}^{(es)}\ + \ E_0^{\rm(3-center)}\,]\,{(\Psi_0^{\rm(3-center)})}^{-1} \ .
\end{equation}

From the viewpoint of the Born-Oppenheimer approximation, widely used in molecular physics, where the particle $m_4=m$ can be associated with an electron and the whole system with a one-electron triatomic molecule, the Hamiltonian ${\cal H}^{\rm(3-center)}$ (\ref{H3cc}) has the meaning of the so-called electronic Hamiltonian. It describes the electronic degrees of freedom of such a molecular system. The ground state energy $E_0^{\rm(3-center)}$ (\ref{e03-3cc}) is called the {\it ground state energy potential curve} or, simply speaking, the potential curve.

The spectrum of the Hamiltonian (\ref{H3cc}) is
\begin{equation}
\label{Ek-3center}
  E^{\rm(3-center)}_{k_1,k_2,k_3} \ = \ E_0^{\rm(3-center)} + \varepsilon_{k_1,k_2,k_3} \ .
\end{equation}

\subsubsection{Laplace-Beltrami operator: underlying geometry}

For the 3-center case, $m_1,m_2,m_3\rightarrow \infty$, the co-metric defined by the coefficients in front of second derivatives in (\ref{algebraic-3cc})

\begin{equation}
\label{gmn33-3cc}
 g_{(\rm 3-center)}^{\mu \nu}(\rho)\ = \left(
 \begin{array}{ccc}
 2\,\frac{\rho_{14}}{m} & \ \frac{\rho_{14} + \rho_{24} - \rho_{12}}{m} & \ \frac{\rho_{14} + \rho_{34} - \rho_{13}}{m} \\
            &                                   &                                   \\
 \frac{\rho_{14} + \rho_{24} - \rho_{12}}{m} & \  2\,\frac{\rho_{24}}{m} & \ \frac{\rho_{24} + \rho_{34} - \rho_{23}}{m} \\
            &  \                                  &                                   \\
 \frac{\rho_{14} + \rho_{34} - \rho_{13}}{m} & \ \frac{\rho_{24} + \rho_{34} - \rho_{23}}{m}  & 2\,\frac{\rho_{34}}{m}
 \end{array}
               \right)  \ ,
\end{equation}

is proportional to $1/m$ and its determinant is given by
\[
 D_{\rm(3-center)}\ \equiv \ {\rm Det} g_{(\rm 3-center)}^{\mu \nu}\ = \ \frac{288}{m^3}\,V_4^2 \ .
\]
We emphasize that the operator (\ref{algebraic-3cc}) is not six but three-dimensional, $\rho_{12}$, $\rho_{13}$ and $\rho_{23}$ are classical variables, i.e. not dynamical. They enter as external parameters only.

Finally, making the gauge transformation
\begin{equation}
\begin{aligned}
         & \Gamma_{(\rm 3-center)}^{-1}\,\Psi_0^{(\rm 3-center)}\, [\,h_{\rm 3-center}^{(es)}\ + \ E_0^{(\rm 3-center)}\,]\,{(\Psi_0^{(\rm 3-center)})}^{-1}\,\Gamma_{(\rm 3-center)} \\ & \
         \ = \ \Gamma_{(\rm 3-center)}^{-1}\, (\,-\Delta_{\rm rad}^{(\rm 3-center)}\,)\,\Gamma_{(\rm 3-center)}
          \ = \  -\Delta^{(\rm 3-center)}_{LB}(\rho) \ + \ V^{(\rm 3-center)}_{\rm eff} \ ,
\end{aligned}
\end{equation}
with the factor,
\begin{equation}
         \Gamma_{(\rm 3-center)} \ = \  V_4^{\frac{3-d}{2}} \quad ,
\end{equation}
we obtain the Laplace-Beltrami operator $\Delta^{(\rm 3-center)}_{LB}(\rho)$, with metric (\ref{gmn33-3cc}), plus the effective potential
\begin{equation}
V^{(\rm 3-center)}_{\rm eff} \ \equiv \   \frac{(d-5)(d-3)\,S^2_{123}}{72\,m\,V_4^2}     \ ,
\end{equation}
where $ S^2_{123}$ is the square of the area of the triangle formed by the three centers, namely the triangle with sizes $\sqrt{\rho_{12}},\sqrt{\rho_{13}}$ and $\sqrt{\rho_{23}}$.

Hence, from the original Hamiltonian (\ref{Hgen}) we arrive at the spectral problem for the Schr\"odinger operator
\begin{equation}\label{}
  H^{(\rm 3-center)}_{LB} \ = \  -\Delta^{(\rm 3-center)}_{LB} \ + \ V^{(\rm 3-center)}_{\rm eff} \ + \ V \ ,
\end{equation}
in the $\rho$-space with $d > 2$.

\clearpage

\section{Symmetry analysis of the four-body problem: arbitrary mass case}
\label{symmetryanalysis}
Here we present the 1st and 2nd order symmetries of the 4-body free Hamiltonian in $S$-state for the case of arbitrary masses, i.e. the kinetic energy $\Delta_{\rm rad}$ (\ref{addition3-3r-M}). The operator $\Delta_{\rm rad}$ is invariant under the ${\mathcal{S}}_3-$group action (permutation between any pair of particles). There exist three 1st order symmetries $J_i$, $i=1,2,3$, linear in $\rho-$variables
\small{
\begin{equation}\label{}
\begin{aligned}
\alpha_1\,J_1 \ & = \ \bigg[   m_1 m_3 \left(\rho _{12}+\rho _{13}-\rho _{23}\right)-m_2 m_3 \left(\rho _{12}-\rho _{13}+\rho _{23}\right)+m_1 m_4 \left(\rho _{12}+\rho _{14}-\rho _{24}\right)
\\ &
-m_2 m_4 \left(\rho _{12}-\rho _{14}+\rho _{24}\right) \bigg]\,\partial_{\rho_{12}}  \ + \ \bigg[ m_2 m_3 \left(\rho _{13}-\rho _{12}+\rho _{23}\right) -m_1 m_2 \left(\rho _{12}+\rho _{13}-\rho _{23}\right)
\\ &
+m_2 m_4 \left(\rho _{14}-\rho _{12}+\rho _{23}-\rho _{34}\right) \bigg]\,\partial_{\rho_{13}}  \ + \ \bigg[ m_2 m_4 \left(\rho _{14}-\rho _{12}+\rho _{24}\right)-m_1 m_2 \left(\rho _{12}+\rho _{14}-\rho _{24}\right)
\\ &
+m_2 m_3 \left(\rho _{13}-\rho _{12}+\rho _{24}-\rho _{34}\right) \bigg]\,\partial_{\rho_{14}} \ + \ \bigg[ m_1 m_2 \rho _{12}-m_1 m_2 \rho _{13}+m_1 m_3 \left(\rho _{12}-\rho _{13}-\rho _{23}\right)
\\ &
+m_1 m_2 \rho _{23}+m_1 m_4 \left(\rho _{12}-\rho _{13}-\rho _{24}+\rho _{34}\right)\bigg]\,\partial_{\rho_{23}}
\ + \  \bigg[ m_1 m_2 \rho _{12}-m_1 m_2 \rho _{14}
\\ &
+m_1 m_4 \left(\rho _{12}-\rho _{14}-\rho _{24}\right)+m_1 m_2 \rho _{24}+m_1 m_3 \left(\rho _{12}-\rho _{14}-\rho _{23}+\rho _{34}\right) \bigg] \,\partial_{\rho_{24}}
\end{aligned}
\end{equation}
}
{\small
\begin{equation}\label{}
\begin{aligned}
\alpha_2\,J_2 \ & = \ \bigg[ m_1 m_3 m_4 \left(\rho _{13}-\rho _{14}-\rho _{23}+\rho _{24}\right)+m_2 m_3 m_4 \left(\rho _{13}-\rho _{14}-\rho _{23}+\rho _{24}\right)
\\ &
+m_3 m_4 \left(m_3+m_4\right) \left(\rho _{13}-\rho _{14}-\rho _{23}+\rho _{24}\right) \bigg]\,\partial_{\rho_{12}} \ + \ \bigg[m_2 m_4^2 \rho _{14} -m_2 m_4^2 \rho _{12}+m_2 m_4^2 \rho _{23}
\\ &
-m_2 m_4^2 \rho _{34}-m_1 m_2 m_4 \left(\rho _{12}+\rho _{13}-\rho _{23}\right)+m_2 m_3 m_4 \left(\rho _{13}-\rho _{12}+\rho _{23}\right)  \bigg]\,\partial_{\rho_{13}} \ + \
\\ &
\bigg[ m_2 m_3^2 \left(\rho _{12}-\rho _{13}-\rho _{24}+\rho _{34}\right)+m_2 m_4 m_3 \left(\rho _{12}-\rho _{14}-\rho _{24}\right)+m_1 m_2 m_3 \left(\rho _{12}+\rho _{14}-\rho _{24}\right)  \bigg]\,\partial_{\rho_{14}}\ + \
\\ &
  \bigg[  m_2 m_4^2 \rho _{23}+m_2 m_4^2 \rho _{24}-m_2 m_4^2 \rho _{34}+m_2 m_3 m_4 \rho _{23}+m_1 m_2 m_4 \left(\rho _{12}-\rho _{13}+\rho _{23}\right)+m_2 m_3 m_4 \rho _{24}
\\ &
-m_2 m_3 m_4 \rho _{34}-m_1 m_3 m_4 \left(\rho _{23}-\rho _{24}+\rho _{34}\right)-m_3 \left(m_3+m_4\right) m_4 \left(\rho _{23}-\rho _{24}+\rho _{34}\right) \bigg]\,\partial_{\rho_{23}} \ + \
\\ &
\bigg[ -m_2 m_3^2 \left(\rho _{23}+\rho _{24}-\rho _{34}\right)-m_2 m_4 m_3 \rho _{23}-m_2 m_4 m_3 \rho _{24}-m_1 m_2 m_3 \left(\rho _{12}-\rho _{14}+\rho _{24}\right)
\\ &
+m_2 m_4 m_3 \rho _{34}+m_1 m_4 m_3 \left(\rho _{24}-\rho _{23}+\rho _{34}\right)+m_4 \left(m_3+m_4\right) m_3 \left(\rho _{24}-\rho _{23}+\rho _{34}\right)  \bigg]\,\partial_{\rho_{24}} \ + \
\\ &
\bigg[  m_2 m_3^2 \left(\rho _{23}-\rho _{24}+\rho _{34}\right)+2 m_2 m_4 m_3 \rho _{23}+m_1 m_2 m_3 \left(\rho _{14}-\rho _{13}+\rho _{23}-\rho _{24}\right)-2 m_2 m_4 m_3 \rho _{24}
\\ &
+m_2 m_4^2 \rho _{23}+m_1 m_2 m_4 \left(\rho _{14}-\rho _{13}+\rho _{23}-\rho _{24}\right)-m_2 m_4^2 \rho _{24}-m_2 m_4^2 \rho _{34} \bigg]\,\partial_{\rho_{34}}
\end{aligned}
\end{equation}
}

\begin{equation}\label{}
\begin{aligned}
& -\alpha_3\,J_3 \ = \ \bigg[ m_3 m_4 \left(\rho _{13}-\rho _{14}-\rho _{23}+\rho _{24}\right) \bigg]\,\partial_{\rho_{12}} \ + \  \bigg[ m_3 m_4 \left(\rho _{13}-\rho _{14}\right)-m_1 m_4 \left(\rho _{13}+\rho _{14}-\rho _{34}\right)
\\ &
+m_3 m_4 \rho _{34} \bigg]\,\partial_{\rho_{13}}\ + \  \bigg[ m_1 m_3 \rho _{13}+m_3 m_4 \left(\rho _{13}-\rho _{14}\right)+m_1 m_3 \rho _{14}
-m_1 m_3 \rho _{34}-m_3 m_4 \rho _{34} \bigg]\,\partial_{\rho_{14}}\ + \
\\ &
\bigg[ m_1 m_4 \left(\rho _{12}-\rho _{13}-\rho _{24}+\rho _{34}\right) \bigg]\,\partial_{\rho_{23}}\ + \ \bigg[ m_1 m_3 \rho _{14}-m_1 m_3 \rho _{12}
+m_1 m_3 \rho _{23}-m_1 m_3 \rho _{34} \bigg]\,\partial_{\rho_{24}}\ + \
\\ &
 \bigg[ m_1 m_3 \rho _{14}-m_1 m_3 \rho _{13}-m_1 m_3 \rho _{34} +m_1 m_4 \left(\rho _{14}-\rho _{13}+\rho _{34}\right) \bigg]\,\partial_{\rho_{34}}
\end{aligned}
\end{equation}
where
\begin{equation}\label{}
\begin{aligned}
& \alpha_1 \ \equiv \ \sqrt{m_1m_2(m_3+m_4)(m_1+m_2+m_3+m_4)}
\\ &
\alpha_2 \ \equiv \  \sqrt{m_2 m_3 m_4 (m_3+m_4)\left(m_1+m_3+m_4\right) \left(m_1+m_2+m_3+m_4\right)}
\\ &
\alpha_3 \ \equiv \  \sqrt{m_1 m_3 m_4 \left(m_1+m_3+m_4\right) } \ ,
\end{aligned}
\end{equation}

$[\, \Delta_{\rm rad},\,J_{i}  \,]=0$, $i=1,2,3$. The operators $J_1,J_2,J_3$ obey the ${\rm so}(3)$ commutation relations

\begin{equation}\label{}
  [\, J_1,\,J_2 \,] \ = \ J_3 \qquad ; \qquad [\, J_3,\,J_1 \,] \ = \ J_2 \qquad ; \qquad[\, J_2,\,J_3 \,] \ = \ J_1 \ .
\end{equation}

Now, there are six 2nd order symmetries linear in the $\rho-$coordinates: $S_1,S_2,S_3,S_4,S_5,S_6$ with Lie commutators $[\,S_j,\,S_k\,]=0$, $1\le j,k\le 6$. Thus, the space is 6-dimensional. The ${\rm so}(3)$ algebra of 1st order symmetries acts on this space and
decomposes it into 2 irreducible pieces: a 1-dimensional piece ($\Delta_{\rm rad}$) and a 5-dimensional piece.
However, there exists a simpler construction of a basis that doesn't make use of the ${\rm so}(3)$ action and gives a basis that is mass independent. The 1st 4 basis vectors $S_1,S_2,S_3,S_4$ are defined by

\begin{equation}\label{}
  \Delta_{\rm rad} \ = \ \frac{1}{m_1}S_1 \ +  \ \frac{1}{m_2}S_2 \ +  \ \frac{1}{m_3}S_3 \ +  \ \frac{1}{m_4}S_4 \ ,
\end{equation}
namely

\begin{equation}\label{}
\begin{aligned}
& S_1 \ = \ -2\bigg( \frac{1}{\mu_{12}} \rho_{12} \,\pa^2_{\rho_{12}} + \frac{1}{\mu_{13}}\rho_{13}\, \pa^2_{\rho_{13}} +\frac{1}{\mu_{14}}\rho_{14}\, \pa^2_{\rho_{14}}  \ + \ (\rho_{12} + \rho_{13} - \rho_{23})\pa_{\rho_{12}}\pa_{\rho_{13}} \ + \
\\ &
\ (\rho_{12} + \rho_{14} - \rho_{24})\pa_{\rho_{12}}\pa_{\rho_{14}} \ + \ (\rho_{13} + \rho_{14} - \rho_{34})\pa_{\rho_{13}}\pa_{\rho_{14}} \bigg) \ - \ d\,( \, \rho_{12} \ + \  \rho_{13} \ + \  \rho_{14}\,)
\\ &
S_2 \ = \ -2\bigg( \frac{1}{\mu_{12}} \rho_{12} \,\pa^2_{\rho_{12}} + \frac{1}{\mu_{23}}\rho_{23}\, \pa^2_{\rho_{23}} +\frac{1}{\mu_{24}}\rho_{24}\, \pa^2_{\rho_{24}}  \ + \ (\rho_{12} + \rho_{23} - \rho_{13})\pa_{\rho_{12}}\pa_{\rho_{23}} \ + \
\\ &
\ (\rho_{12} + \rho_{24} - \rho_{14})\pa_{\rho_{12}}\pa_{\rho_{24}} \ + \ (\rho_{23} + \rho_{24} - \rho_{34})\pa_{\rho_{23}}\pa_{\rho_{24}} \bigg) \ - \ d\,( \, \rho_{12} \ + \  \rho_{23} \ + \  \rho_{24}\,)
\\ &
S_3 \ = \ -2\bigg( \frac{1}{\mu_{13}} \rho_{13} \,\pa^2_{\rho_{13}} + \frac{1}{\mu_{23}}\rho_{23}\, \pa^2_{\rho_{23}} +\frac{1}{\mu_{34}}\rho_{34}\, \pa^2_{\rho_{34}}  \ + \ (\rho_{13} + \rho_{23} - \rho_{12})\pa_{\rho_{13}}\pa_{\rho_{23}} \ + \
\\ &
\ (\rho_{23} + \rho_{34} - \rho_{24})\pa_{\rho_{23}}\pa_{\rho_{34}} \ + \ (\rho_{13} + \rho_{34} - \rho_{14})\pa_{\rho_{13}}\pa_{\rho_{34}} \bigg) \ - \ d\,( \, \rho_{13} \ + \  \rho_{23} \ + \  \rho_{34}\,)
\\ &
S_4 \ = \ -2\bigg( \frac{1}{\mu_{14}} \rho_{14} \,\pa^2_{\rho_{14}} + \frac{1}{\mu_{24}}\rho_{24}\, \pa^2_{\rho_{24}} +\frac{1}{\mu_{34}}\rho_{34}\, \pa^2_{\rho_{34}}  \ + \ (\rho_{24} + \rho_{34} - \rho_{23})\pa_{\rho_{24}}\pa_{\rho_{34}} \ + \
\\ &
\ (\rho_{14} + \rho_{24} - \rho_{12})\pa_{\rho_{14}}\pa_{\rho_{24}} \ + \ (\rho_{14} + \rho_{34} - \rho_{13})\pa_{\rho_{14}}\pa_{\rho_{34}} \bigg) \ - \ d\,( \, \rho_{14} \ + \  \rho_{24} \ + \  \rho_{34}\,) \ ,
\end{aligned}
\end{equation}

whilst $S_5$ and $S_6$ are determined by the commutator $ [\, J_3,\,S_1  \,]$ as follows

\begin{equation}\label{}
  \frac{1}{2}\sqrt{\frac{m_1 m_3 m_4 \left(m_1+m_3+m_4\right)}{m_1}}[\, J_3,\,S_1  \,] \ = \ m_3\,S_5 \ - \ m_4\,S_6 \ .
\end{equation}

Explicitly, they are given by

\begin{equation}\label{}
\begin{aligned}
& S_5 \ = \  2\,\rho_{14} \,\pa^2_{\rho_{14}} \ + \ (\rho _{12}+\rho _{14}-\rho _{24})\pa_{\rho_{12}}\pa_{\rho_{14}} \ - \
 (\rho _{12}-\rho _{14}+\rho _{24})\pa_{\rho_{12}}\pa_{\rho_{24}}
\\ &
\ + \ (-\rho _{13}+\rho _{14}+\rho _{23}-\rho _{24})\pa_{\rho_{12}}\pa_{\rho_{34}} \ + \ (\rho _{13}+\rho _{14}-\rho _{34})\pa_{\rho_{13}}\pa_{\rho_{14}} \ + \ (-\rho _{12}+\rho _{14}+\rho _{23}-\rho _{34})\pa_{\rho_{13}}\pa_{\rho_{24}}
\\ &
 \ - \ (\rho _{13}-\rho _{14}+\rho _{34})\pa_{\rho_{13}}\pa_{\rho_{34}} \ + \ (-\rho _{12}+\rho _{14}+\rho _{24})\pa_{\rho_{14}}\pa_{\rho_{24}} \ + \ (-\rho _{13}+\rho _{14}+\rho _{34})\pa_{\rho_{14}}\pa_{\rho_{34}} \ + \ d\,\pa_{\rho_{14}}
\\ &
S_6 \ = \  2\,\rho_{13} \,\pa^2_{\rho_{13}} \ + \ (\rho _{12}+\rho _{13}-\rho _{23})\pa_{\rho_{12}}\pa_{\rho_{13}} \ - \ (\rho _{12}-\rho _{13}+\rho _{23})\pa_{\rho_{12}}\pa_{\rho_{23}}
\\ &
 \ + \ (\rho _{13}-\rho _{14}-\rho _{23}+\rho _{24})\pa_{\rho_{12}}\pa_{\rho_{34}} \ + \ (\rho _{13}+\rho _{14}-\rho _{34})\pa_{\rho_{13}}\pa_{\rho_{14}} \ + \ (-\rho _{12}+\rho _{13}+\rho _{23})\pa_{\rho_{13}}\pa_{\rho_{23}}
\\ &
  \ + \ (\rho _{13}-\rho _{14}+\rho _{34})\pa_{\rho_{13}}\pa_{\rho_{34}} \ + \ (-\rho _{12}+\rho _{13}+\rho _{24}-\rho _{34})\pa_{\rho_{14}}\pa_{\rho_{23}} \ + \ (\rho _{13}-\rho _{14}-\rho _{34})\pa_{\rho_{14}}\pa_{\rho_{34}}  \ + \ d\,\pa_{\rho_{13}} \ .
\end{aligned}
\end{equation}

One can verify directly that $[\, \Delta_{\rm rad},\,S_i  \,]=0$, $i=1,2,\ldots,6$. Furthermore the set $(S_1,S_2,\ldots,S_6)$ is linearly independent. Hence, it follows that the system is integrable. As for the 2nd order symmetry algebra of $\Delta_{\rm rad}$, quadratic in the $\rho-$variables, the dimension of the space is $21$ and under the ${\rm so}(3)$ action it splits into five irreducible sub-spaces, two of dimension 1 ($\ell=0$), two of dimension 5 ($\ell=2$) and one of dimension 9. In fact, it can be shown that the free Hamiltonian $\Delta_{\rm rad}$ is maximally superintegrable.

\newpage

\section{Towards the reduction to the planar case $d=2$}
\label{reduction}

At $d=2$ (planar systems) the number of independent $\rho$-variables reduces from 6 to 5 and the expression (\ref{addition3-3r-M}) for the operator $\Delta_{\rm rad}$ ceases to be valid. In particular, the determinant of the metric defined by the coefficients of the 2nd order derivatives in (\ref{gmn33-rho-det-M}) vanishes. This makes the case $d=2$ quite distinct from $d\geq3$. We mention that the limit $d \rightarrow 2$ for the case of identical masses $m_1=m_2=m_3=m_4=1$ was already treated in \cite{MTE:2018}. Here we consider the most general case of arbitrary masses.

For $d = 2$ the volume $V_4$ (\ref{V4}) of the tetrahedron of interaction vanishes. Effectively, it leads to a reduction of the
dimension of the space of relative distances from 6 to 5. In order to see this dimensional reduction explicitly, let us change
variables
\begin{equation}\label{}
  (\, \rho_{12},\,\rho_{13},\,\rho_{14},\,\rho_{23},\,\rho_{24},\,\rho_{34}  \,) \rightarrow (\, \rho_{12},\,\rho_{13},\,\rho_{14},\,\rho_{23},\,\rho_{24},\,{{\cal V} \equiv V_4^2}  \,)
\end{equation}
in the free Hamiltonian $\Delta_{\rm rad}$ (\ref{addition3-3r-M}). i.e. we replace
\[
\rho_{34}\ \rightarrow \ {\cal V}
\]
and consider ${\cal V}$ as new dynamical variable. Then, it follows that

{\small
\begin{equation}
\begin{aligned}
\label{deltad2}
& \Delta_{\rm rad}\ =  \  \frac{{S}\,{\cal V}}{72\,m_1\,m_2\,m_3\,m_4}\pa^2_{\cal V} \ + \  2\bigg( \frac{1}{\mu_{12}} \rho_{12} \,\pa^2_{\rho_{12}} + \frac{1}{\mu_{13}}\rho_{13}\, \pa^2_{\rho_{13}} +\frac{1}{\mu_{14}}\rho_{14}\, \pa^2_{\rho_{14}}
 +\frac{1}{\mu_{23}}\rho_{23}\, \pa^2_{\rho_{23}} +\frac{1}{\mu_{24}}\rho_{24}\, \pa^2_{\rho_{24}}
 \bigg)
\\ &
 +  \frac{2}{m_1} \bigg({(\rho_{12} + \rho_{13} - \rho_{23})}\pa_{\rho_{12}}\pa_{\rho_{13}}\ +
          {(\rho_{12} + \rho_{14} - \rho_{24})}\pa_{\rho_{12}}\pa_{\rho_{14}}\ +
          {(\rho_{13} + \rho_{14} - \rho_{34})}\pa_{\rho_{13}}\pa_{\rho_{14}} \bigg)
\\ &
+  \frac{2}{m_2} \bigg((\rho_{12} + \rho_{23} - \rho_{13})\pa_{\rho_{12}}\pa_{\rho_{23}}\ +
          (\rho_{12} + \rho_{24} - \rho_{14})\pa_{\rho_{12}}\pa_{\rho_{24}}\ +
          (\rho_{23} + \rho_{24} - \rho_{34})\pa_{\rho_{23}}\pa_{\rho_{24}}
    \bigg)
\\ &
+  \frac{2}{m_3} (\rho_{13} + \rho_{23} - \rho_{12})\pa_{\rho_{13}}\pa_{\rho_{23}}
\ + \   \frac{2}{m_4} (\rho_{14} + \rho_{24} - \rho_{12})\pa_{\rho_{14}}\pa_{\rho_{24}}
\\ &
+  (4\,{\cal V}\, \pa_{\cal V} \,+\,d)\,\bigg(\frac{1}{\mu_{12}}\pa_{\rho_{12}} + \frac{1}{\mu_{13}}\pa_{\rho_{13}}+ \frac{1}{\mu_{14}}\pa_{\rho_{14}}+ \frac{1}{\mu_{23}}\pa_{\rho_{23}}+ \frac{1}{\mu_{24}}\pa_{\rho_{24}}\bigg) \ + \ \frac{(d-2){ S}}{144\,m_1\,m_2\,m_3\,m_4} \pa_{\cal V}       \ ,
\end{aligned}
\end{equation}
}
where
\begin{equation}\label{vaS}
\begin{aligned}
  {S} \ & = \   m_1\,m_2\,m_3\,(2\,\rho_{12}\,\rho_{23}+2\,\rho_{13}\,\rho_{23}+2\,\rho_{12}\,\rho_{13} -\rho_{12}^2-\rho_{13}^2-\rho_{23}^2) \ + \  m_1\,m_2\,m_4\,(2\,\rho_{12}\,\rho_{24}+2\,\rho_{14}\,\rho_{24}
\\ &  \ +  2\,\rho_{12}\,\rho_{14} -\rho_{12}^2-\rho_{14}^2-\rho_{24}^2)
\ + \ m_1\,m_3\,m_4\,(2\,\rho_{13}\,\rho_{34}+2\,\rho_{14}\,\rho_{34}+2\,\rho_{13}\,\rho_{14} -\rho_{13}^2-\rho_{14}^2-\rho_{34}^2)
\\ &
\ + \ m_2\,m_3\,m_4\,(2\,\rho_{23}\,\rho_{24}+2\,\rho_{23}\,\rho_{34}+2\,\rho_{24}\,\rho_{34} -\rho_{23}^2-\rho_{24}^2-\rho_{34}^2) \ ,
\end{aligned}
\end{equation}
(c.f.\ref{Sweighted}) is the weighted sum of the four areas (squared) of the faces of the tetrahedron of interaction.

The variable $\rho_{34}$ appearing in the above formulas is not an independent coordinate, it is given by
\begin{equation}\label{rho34}
\begin{aligned}
  \rho_{34} \ & = \ -\frac{\rho _{12}^2-\left(\rho _{13}+\rho _{14}+\rho _{23}+\rho _{24}\right) \rho _{12}+\left(\rho _{13}-\rho _{23}\right) \left(\rho _{14}-\rho _{24}\right)}{2\, \rho _{12}}
\\ &
\pm\,\frac{1}{2\, \rho _{12}}\sqrt{\left(\rho _{12}^2-2 \left(\rho _{13}+\rho _{23}\right) \rho _{12}+\left(\rho _{13}-\rho _{23}\right){}^2\right) \left(\rho _{12}^2-2 \left(\rho _{14}+\rho _{24}\right) \rho _{12}+\left(\rho _{14}-\rho _{24}\right){}^2\right)\,-\,576\, \rho _{12}\, {\cal V}} \ ,
\end{aligned}
\end{equation}

therefore, the operator (\ref{deltad2}) is not algebraic.

At $d=2$ (thus, ${\cal V} \equiv 0 $), from (\ref{deltad2}) we obtain the five-dimensional operator
{\small
\begin{equation}
\begin{aligned}
\label{Deld2}
& \Delta_{\rm rad}^{(d=2)}\mid_{{\cal V} = 0} \ =  \   2\bigg( \frac{1}{\mu_{12}} \rho_{12} \,\pa^2_{\rho_{12}} + \frac{1}{\mu_{13}}\rho_{13}\, \pa^2_{\rho_{13}} +\frac{1}{\mu_{14}}\rho_{14}\, \pa^2_{\rho_{14}}
 +\frac{1}{\mu_{23}}\rho_{23}\, \pa^2_{\rho_{23}} +\frac{1}{\mu_{24}}\rho_{24}\, \pa^2_{\rho_{24}}
 \bigg)
\\ &
 +  \frac{2}{m_1} \bigg({(\rho_{12} + \rho_{13} - \rho_{23})}\pa_{\rho_{12}}\pa_{\rho_{13}}\ +
          {(\rho_{12} + \rho_{14} - \rho_{24})}\pa_{\rho_{12}}\pa_{\rho_{14}}\ +
          {(\rho_{13} + \rho_{14} - \rho_{34})}\pa_{\rho_{13}}\pa_{\rho_{14}} \bigg)
\\ &
+  \frac{2}{m_2} \bigg((\rho_{12} + \rho_{23} - \rho_{13})\pa_{\rho_{12}}\pa_{\rho_{23}}\ +
          (\rho_{12} + \rho_{24} - \rho_{14})\pa_{\rho_{12}}\pa_{\rho_{24}}\ +
          (\rho_{23} + \rho_{24} - \rho_{34})\pa_{\rho_{23}}\pa_{\rho_{24}}
    \bigg)
\\ &
+  \frac{2}{m_3} (\rho_{13} + \rho_{23} - \rho_{12})\pa_{\rho_{13}}\pa_{\rho_{23}}
\ + \   \frac{2}{m_4} (\rho_{14} + \rho_{24} - \rho_{12})\pa_{\rho_{14}}\pa_{\rho_{24}}
\\ &
+  2\,\bigg(\frac{1}{\mu_{12}}\pa_{\rho_{12}} + \frac{1}{\mu_{13}}\pa_{\rho_{13}}+ \frac{1}{\mu_{14}}\pa_{\rho_{14}}+ \frac{1}{\mu_{23}}\pa_{\rho_{23}}+ \frac{1}{\mu_{24}}\pa_{\rho_{24}}\bigg)     \ .
\end{aligned}
\end{equation}
}

Due to the relation (\ref{rho34}) this operator (\ref{Deld2}) is not algebraic either. We were not able to find new five variables for which (\ref{Deld2}) becomes algebraic in full generality. Nevertheless, in (\ref{Deld2}) a sub-structure remains algebraic in the limits $d=1,2$ as described below.

\subsection{The case $d\geq2$: $(P,S)-$representation}

The operator $\Delta_{\rm rad}$ (\ref{addition3-3r-M}) can be further decomposed into the sum of 2 operators
\begin{equation}
\label{DEsum0}
\Delta_{\rm rad} \ = \ \Delta_{P,S} \ + \   \Delta_{w} \ ,
\end{equation}
with the following properties:
\begin{itemize}
  \item $\Delta_{P,S}=\Delta_{P,S}(P,S)$ is an algebraic operator for any $d>1$. It depends on the \emph{volume variables} $S$ (\ref{vaS}) and
\begin{equation}\label{}
  P \ \equiv \ \frac{m_1 m_2 \,\rho_{12}  \,+\,m_1 m_3 \,\rho_{13} \,+\,m_1 m_4 \,\rho_{14} \,+\,m_2 m_3 \,\rho_{23}\, +\,m_2 m_4 \,\rho_{24} \,+\,m_3 m_4 \,\rho_{34} }{m_1\,+\,m_2\,+\,m_3\,+\,m_4} \ ,
\end{equation}
(c.f.\ref{Pweighted}) and its derivatives alone. Explicitly,
\begin{equation}
\begin{aligned}
{\label{DEg0}}
\Delta_{P,S} \ = \  2\,P\,\pa^2_{P,P}\  + \ 8\,M\,P\,S\,\pa^2_{S,S}   \  + \ 8\,M\,(d-1)\,P\,\pa_{S}  \  + \ 3\,d\,\pa_{P}      \ ,
\end{aligned}
\end{equation}
$M=m_1 + m_2 + m_3 + m_4$.
  \item $\Delta_{w}=\Delta_{w}(P,S,w)$ depends on $P$, $S$ and $w$ where the latter denotes any set of variables such that the transformation $\{\rho_{ij}\}\rightarrow\{P,S,w\}$ is not singular. This operator annihilates any $(P,S)$-dependent function, namely $\Delta_{w}\,f(P,S)\ =\ 0$\,. We do not give its explicit form.
  \item $[\Delta_{P,S},\,\Delta_{w}] \ \neq 0 $\ .
\end{itemize}

The decomposition (\ref{DEsum0}) implies that for the infinite family of two-parametric potentials
\begin{equation}\label{}
  V \ = \  V(P,S) \ ,
\end{equation}
the spectral problem for the radial operator
\begin{equation}\label{}
  H_{\rm rad}\ \equiv\ -\De_{\rm rad} \ + \  V \ ,
\end{equation}
is reduced to the simpler two-dimensional eigenvalue equation
\begin{equation}\label{HPd0}
  H_{P,S}\,\psi \ \equiv\ [\,-\De_{P,S} \ + \  V(P,S)\,]\,\psi \ = \ \epsilon \, \psi \ .
\end{equation}

We call (\ref{HPd0}) the $(P,S)-$representation. We emphasize that the nature of the volume variables $P$ and $S$ is purely geometrical. Putting $m_1=m_2=m_3=m_4=1$ and replacing $P\rightarrow \frac{P}{4}$ , $S\rightarrow 16\,S$ we obtain the results presented in \cite{MTE:2018} for the equal-mass case. 

\subsection{The case $d\geq1$: $P-$representation}

Clearly, the operator $\Delta_{\rm rad}$ (\ref{addition3-3r-M}) splits into the sum of 2 operators
\begin{equation}
\label{DEsum}
\Delta_{\rm rad} \ = \ \Delta_{P} \ + \   \Delta_{q} \ ,
\end{equation}
with the following properties:
\begin{itemize}
  \item $\Delta_{P}=\Delta_{P}(P)$ is an algebraic operator for any $d\geq 1$. It depends only on the \emph{volume variable} 
\begin{equation}\label{}
  P \ = \ \frac{m_1 m_2 \,\rho_{12}  \,+\,m_1 m_3 \,\rho_{13} \,+\,m_1 m_4 \,\rho_{14} \,+\,m_2 m_3 \,\rho_{23}\, +\,m_2 m_4 \,\rho_{24} \,+\,m_3 m_4 \,\rho_{34} }{m_1\,+\,m_2\,+\,m_3\,+\,m_4} \ ,
\end{equation}
(c.f.\ref{Pweighted}) and its derivatives. Explicitly,
\begin{equation}
\begin{aligned}
{\label{DEg}}
\Delta_{P} \ = \  2\,P\,\pa^2_{P,P}\  + \ 3\,d\,\pa_{P}      \ .
\end{aligned}
\end{equation}

  \item $\Delta_{q}=\Delta_{q}(P,q)$ depends on $P$ and $q$ where the latter denotes any set of variables such that the transformation $\{\rho_{ij}\}\rightarrow\{P,q\}$ is not singular. This operator annihilates any $P$-dependent function, namely $\Delta_{q}\,f(P)\ =\ 0$\,. We do not give its explicit form.
  \item $[\Delta_{P},\,\Delta_{q}] \ \neq 0 $\ .
\end{itemize}

The decomposition (\ref{DEsum}) implies that for the infinite family of one-parametric potentials
\begin{equation}\label{}
  V \ = \  V(P) \ ,
\end{equation}
the spectral problem for the radial operator
\begin{equation}\label{}
  H_{\rm rad}\ \equiv\ -\De_{\rm rad} \ + \  V \ ,
\end{equation}
is reduced to the one-dimensional eigenvalue equation
\begin{equation}\label{HPd}
  H_{P}\,\psi \ \equiv\ [\,-\De_{P} \ + \  V(P)\,]\,\psi \ = \ \epsilon \, \psi \ ,
\end{equation}
which we call the $P-$representation. 

\textbf{Remarks.} It is worth mentioning that the variable $P$ is the moment of inertia (scaled) of the system. Just as for the case of equal masses $m_1=m_2=m_3=m_4$ the variable $P$ becomes proportional to the hyperspherical radius appearing in the hyperspherical-harmonic expansion method \cite{Delves:1958}-\cite{Smith}.

\subsubsection{Sextic anharmonic potential $V^{(qes)}(P)$ : quasi-exact-solvability}

The remarkable property of the operator $H_{P}$ (\ref{HPd}) is its gauge-equivalence to the Schr\"{o}dinger operator.
Making a gauge rotation with the factor
\[
\Gamma_P \ = \ P^{-\frac{1}{4}(3d-1)} \ ,
\]
we obtain that
\begin{equation}
\label{HLBP}
 H_P^{(LB)} \ \equiv \ \Gamma_P^{-1}\,H_P\,\Gamma_P \ = \  -\Delta_{LB}^{(P)} \ + \ V(P) \ + \ U_{\rm eff}(P) \ ,
\end{equation}
where
\begin{equation}\label{}
  U_{\rm eff}(P) \ = \  (d-1) (3 \,d-1) \frac{3}{8 \,P} \ ,
\end{equation}
is an effective potential, and $\Delta_{LB}^{(P)}$ stands for the Laplace-Beltrami operator with $g^{11}=2\,P$, namely
\[
\Delta_{LB}^{(P)} \ = \ 2\,P\,\pa^2_{P,P}\  + \ \pa_{P}  \ .
\]
For the 4-body harmonic oscillator potential
\begin{equation}
\label{VPm}
\begin{aligned}
 V_P \ = & \  V^{(es)} \ \equiv \ 2\,\om^2\,P
  \\ & = \ \frac{2\,\om^2}{M}(m_1 m_2 \,\rho_{12}  \,+\,m_1 m_3 \,\rho_{13} \,+\,m_1 m_4 \,\rho_{14} \,+\,m_2 m_3 \,\rho_{23}\, +\,m_2 m_4 \,\rho_{24} \,+\,m_3 m_4 \,\rho_{34}) \ ,
\end{aligned}
\end{equation}
where $M=m_1+m_2+m_3+m_4$, the eigenvalue equation (\ref{HPd}) has infinitely-many eigenfunctions, see below.

Let us take the function

\begin{equation}
\label{psP}
  \psi_0 \ = \ P^{\frac{1}{4}(3d-1)}\,e^{-\omega\,P-\frac{A}{2}\,P^2} \ ,
\end{equation}

where $A$ is a positive constant, and seek the potential $V_0$ for
which this (\ref{psP}) is the ground state function for the Hamiltonian $H_P^{(LB)}$ (\ref{HLBP}). This potential
can be found immediately by calculating the ratio

\[
\frac{[\, \Delta_{LB}^{(P)}\,\psi_0  \,]}{\psi_0} \ = \ V_0 \ - \ E_0 \ ,
\]

The result is

\begin{equation}
\label{VLBP0}
  V_0 \ = \ \frac{3 (d-1) (3 d-1)}{8 \,P} \ + \ 2 \omega ^2\,P \ + \ A\,\big[\,2 \,A\, P^3 \ + \ 4\,P^2\, \omega \ -  \  P\, (3d+2)\,\big] \ ,
\end{equation}

with the energy of the ground state

\begin{equation}
\label{E0PLB}
  E_0 \ = \ 3\,d\,\omega \ .
\end{equation}

Now, let us take the Hamiltonian $H_P^{(LB)}= -\Delta_{LB}^{(P)} + V_0$ with potential (\ref{VLBP0}), subtract
$E_0$ (\ref{E0PLB}) and make the gauge rotation with $\psi_0$ (\ref{psP}). As the result we obtain the $sl(2,{\bf R})$ Lie algebraic operator $h^{(qes)}_P$ with additional potential $\Delta V_N$

\begin{equation}\label{}
\begin{aligned}
  \psi_0^{-1}\,( -\Delta_{LB}^{(P)} \,+\, V_0\, -\, E_0  )\,\psi_0 \ & =  \ -\Delta_P \ + \ 4\,P\,(A\,P \,+\, \omega)\,\partial_P
\\ & = \ h^{(qes)}_P  \ + \ \Delta V_N  \ ,
\end{aligned}
\end{equation}
where
\[
\Delta V_N  \ = \ 4\,N\,A\,P  \ ,
\]
and $h^{(qes)}_P$ can be expressed in terms of the $sl(2,{\bf R})$ algebra generators
\begin{equation}\label{}
  J^{+}(N) \ = \ P^2\,\partial_P\,-\, N\,P \quad , \quad J^{0}(N) \ = \ 2\,P\,\partial_P-N \quad , \quad J^{-} \ = \ \partial_P \ .
\end{equation}
Explicitly,
\begin{equation}\label{}
 h^{(qes)}_P \ = \ -  J^{-}J^{0} \ - \ (3d+N-2)J^{-}  \ + \  4\,A\,J^+ \ + \ 2\,\omega\,J_0 \ + \ 2\,N\,\omega \ .
\end{equation}
For $A \neq 0$, it is evident that for a positive integer $N$ the operator $h^{(qes)}_P(J)$ has a finite-dimensional invariant subspace
\begin{equation}
\label{}
     {\cal P}_{N}\ =\ \langle P^{p_1} \vert \ 0 \le p_1 \le N \rangle\ ,
\end{equation}
with dim${\cal P}_{N} \sim N$ at large $N$. Eventually, we arrive at the quasi-exactly-solvable
Schr\"{o}dinger operator:
\begin{equation}\label{HPqes}
H_P^{(qes)} \ = \  -\Delta_{LB}^{(P)} + V^{(qes)}(P) \ ,
\end{equation}
where
\begin{equation}\label{}
V^{(qes)} \ = \   \frac{3 (d-1) (3 d-1)}{8 \,P} \ + \ 2 \omega ^2\,P \ + \ A\,\big[\,2 \,A\, P^3 \ + \ 4\,P^2\, \omega \ -  \  P\, (3d+2-4N)\,\big] \ + \ 2\,N\,\omega \ .
\end{equation}
The eigenvalue problem for the operator $H_P^{(qes)}$ (\ref{HPqes}) admits solutions of the factorized form

\begin{equation}\label{}
\psi_N(P) \ = \  {\rm Pol}_N(P) \times \psi_0 \ = \ {\rm Pol}_N(P) \times P^{\frac{1}{4}(3d-1)}\,e^{-\omega\,P-\frac{A}{2}\,P^2}\ ,
\end{equation}
with $\psi_0$ given by (\ref{psP}) and ${\rm Pol}_N(P)$ being a polynomial of degree $N=0,1,2\ldots$ in the variable $P$.

Finally, for $A=0$ the operator $H_P^{(qes)}$ (\ref{HPqes}) becomes exactly-solvable
\begin{equation}\label{HPes}
H_P^{(es)} \ \equiv H_P^{(qes)}\mid_{A=0}  \ = \  -\Delta_{LB}^{(P)} + V^{(es)}(P) \ ,
\end{equation}
here
\begin{equation}\label{}
V^{(es)}  \ \equiv V^{(qes)}\mid_{A=0}  \ = \  \frac{3 (d-1) (3 d-1)}{8 \,P} \ + \ 2 \omega ^2\,P \ + \ 2\,N\,\omega \ .
\end{equation}
In this case, $H_P^{(es)}$ (\ref{HPes}) possesses solutions of the factorized form
\begin{equation}\label{}
\psi_N(P) \ = \  L_N^{(\frac{3d-2}{2})}(2\,P\,\omega) \times P^{\frac{1}{4}(3d-1)}\,e^{-\omega\,P} \ ,
\end{equation}
where $L_N^{(\frac{3d-2}{2})}(2\,P\,\omega)$ is a generalized Laguerre polynomial and the spectra of energies
\begin{equation}\label{}
  \epsilon_N \ = \ (\,3\,d \ + \ 4\,N)\,\omega \ ,
\end{equation}
is equidistant. We again point out that the form of this operator implies the existence of a subfamily of solutions of the original 4-body problem in the space of relative motion, which depend only on the variable $P$\,.

\section{Remarks and Conclusions}
\label{conclusions}

\bigskip

We defined in $\mathbb{R}^d$ ($d>2$) a 4-body harmonic oscillator with pairwise interactions. It is shown that for $S$-states - the states with zero total angular momentum - if the $6$-dimensional space of relative motion is parameterized by relative distances squared $\rho_{ij}=r_{ij}^2$ the problem has a hidden algebra $sl(7, {\bf R})$. It implies that for arbitrary masses and spring constants the system is exactly-solvable. In general, the problem does not allow separation of variables. The corresponding eigenvalues are linear in the six quantum numbers and the eigenfunctions are polynomials in six variables multiplied by Gaussian functions in relative distances. The cases of identical particles as well as the atomic-like and molecular-like cases were analyzed in detail. For any $d\geq 1$, a quasi-exactly solvable model was constructed where the potential is given by a three-parameter cubic polynomial in the variable $P$, hence a sextic potential in the original relative distances $r_{ij}$. At special values of the parameters, the model becomes exactly solvable and corresponds to a certain 4-body oscillator system. The Schr\"{o}dinger operator with hidden algebra $sl(2, {\bf R})$ that describe such $P$-dependable QES eigenfunctions was derived explicitly. By construction these solutions are also eigenfunctions of the original problem. The 1st and 2nd order symmetries of the 4-body free Hamiltonian for $S$-states were presented as well. For specific values of masses the new $3d$ non-conformally flat 4-body oscillator system is superintegrable. Almost all of the structure and classification theory for superintegrable systems applies only to conformally flat spaces, e.g. \cite{KKM:2018}. Examples on non-conformally flat spaces are relatively rare and thus valuable.

The generalization to the $n$-body case in $\mathbb{R}^{d}$ with $d \geq n-1$ is straightforward. In the case $n=5$ (five-body system) with $d\geq 4$, assuming that the potential only depends on the 10 relative distances between particles, the $S-$state solutions are eigenfunctions of the reduced Hamiltonian in the $\rho-$representation

\begin{equation}
\label{redradN5}
  { H}^{(\rm n=5)}_{\rm rad}(\rho) \ =\ -\De^{(\rm n=5)}_{\rm rad}(\rho)\ +\ V(\rho) \ ,
\end{equation}
where $\De^{(\rm n=5)}_{\rm rad}$ is an algebraic operator:
\begin{equation}
\label{eqn20}
\De^{(\rm n=5)}_{\rm rad}(\rho)\ =\
2\,\sum_{i \neq j,i\neq k, j< k}^{5}\,\frac{1}{m_i}(\rho_{ij} + \rho_{ik} - \rho_{jk})\pa_{\rho_{ij}}\pa_{\rho_{ik}}\ + \
2\,\sum_{i<j}^5 \bigg(\frac{m_i+m_j}{m_i m_j} \bigg) \rho_{ij} \pa^2_{\rho_{ij}} +  d\, \sum_{i<j}^5\,\bigg(\frac{m_i+m_j}{m_i m_j}\bigg)\pa_{\rho_{ij}} \ .
\end{equation}

The determinant of the co-metric associated with $\De^{(\rm n=5)}_{\rm rad}$ is proportional to $V_5^2$ (the square of the volume of the 5-vertex polytope of interaction) given by the Cayley-Menger determinant, hence, a homogeneous polynomial of degree 6 in $\rho$-variables. The operator (\ref{eqn20}) is $sl(11,{\bf R})$-Lie algebraic - it can be rewritten in terms of the generators of the maximal affine subalgebra $b_{11}$ of the algebra $sl(11,{\bf R})$, realized by first order differential operators.

The degeneration from $d\geq4$ to the physical three-dimensional space $d=3$ corresponds to the geometric condition $V_5=0$. To study this reduction, in (\ref{eqn20}) we first introduce $V_5$ as one of the dynamical variables, and afterwards, we take the limit $V_5\rightarrow 0$. However, the so obtained nine-dimensional operator in $\rho$-variables is not algebraic anymore. The existence of a set of variables for which this reduced operator becomes algebraic is still an open question.

\section*{Author's contributions}

All authors contributed equally to this work.

\section*{Acknowledgments}

M.A.E.R is thankful to the Centre de Recherches Math\'{e}matiques, Universit\'{e} de Montr\'{e}al, for kind hospitality extended to him where this work was initiated and is grateful to ICN UNAM, Mexico, for the kind hospitality during his numerous visits, where the work was continued. A.V.T. is thankful to University of Minnesota, USA for kind hospitality extended to him where this work was initiated and where it was mostly completed. W.M. was partially supported by a grant from the Simons Foundation (\# 412351 to Willard Miller, Jr.). This work is partially supported by CONACyT Grant No.A1-S-17364 and DGAPA Grant No.IN113819 (Mexico).

\end{document}